# Two-dimensional heat transfer and melt flow in the laser-impact zone at selective laser melting of thin walls – II: Modeling

Andrey V. Gusarov[*], Tatiana V. Tarasova, Sergey N. Grigoriev

*Moscow State University of Technology STANKIN, Vadkovsky Per. 3a, 127055 Moscow, Russia*

**Abstract.** Part II of the present work concerns modeling and analyzing the experimental data obtained in Part I. A computational fluid dynamics (CFD) model of two-dimensional conductive heat transfer and thermocapillary-driven convection is developed. The conservation laws for mass, momentum, and energy are numerically solved by a second-order Godunov finite-volume method using an original Riemann solver developed for the applied equation of state. The CFD model is validated by comparison with the experiments. Formation of two outward vortices in the melt pool is revealed. Surface-active impurities can make the surface tension-temperature function non-monotonous giving raise additional vortices with the opposite inward flow direction. The influence of melt convection on the melt pool size is not considerable in the conditions of selective laser melting (SLM). The flow velocity in the melt pool is around or less than the laser scanning speed. This means insufficient mixing in the melt pool. One cannot expect a complete homogenization of chemical composition when using powder blends in SLM. The correlation between the error of the Rosenthal model and the latent heat of fusion is carefully studied resulting an analytical model for estimating the melt pool depth. In the studied SLM cases, the accuracy of the developed analytical model is within 10% relative the CFD model. The obtained experimental and theoretical results indicate that the melt depth is approximately proportional to linear energy density LED in the typical conditions of SLM. This can be useful when optimizing the SLM process.

______________________________________________________

[*] Corresponding author e-mail address: av.goussarov@gmail.com

**Nomenclature**

| | |
|---|---|
| *A* | absorptance |
| *a* | thermal diffusivity |
| *B* | bulk modulus |
| *b* | dimensionless density jump |
| *C* | specific heat per unit volume |
| *c* | sound speed |
| *D* | melt pool depth |
| *E* | energy per unit volume |
| $\mathcal{E}$ | elastic component of internal energy per unit volume |
| **e** | symmetric component of tensor $\nabla \mathbf{u}$ |
| *e* | Euler number, the base of the natural logarithm |
| *H* | enthalpy per unit volume |
| **I** | second-order identity tensor |
| *J* | Riemann invariant |
| K() | modified Bessel function of the second kind |
| *L* | melt pool length |
| LED | linear energy density |
| $L_m$ | specific enthalpy of melting per unit volume |
| **n** | unit normal to the surface |
| *P* | laser power |
| *p* | pressure |
| **Q** | energy flow |
| **q** | conductive heat flow |
| Pe | thermal Peclet number defined by the flow velocity |
| Re | Reynolds number |
| *r* | radius |
| *s* | square root of density |
| *s*() | level function |
| *T* | temperature |
| $\mathcal{T}$ | thermal component of internal energy per unit volume |
| *t* | time |
| tr() | trace |
| *U* | internal energy per unit volume |
| **u** | flow velocity |
| *V* | scanning velocity |
| *v* | front velocity |
| $W$ | melt pool width |
| $x$, $z$ | Cartesian coordinates |

*Greek Symbols*

| | |
|---|---|
| $\alpha$ | surface tension |
| $\beta$ | temperature coefficient of surface tension |
| $\delta$ | wall thickness |
| $\eta$ | dynamic viscosity |
| $\lambda$ | thermal conductivity |
| $\Pi$ | momentum flow |
| $\Pi$ | thermal Peclet number defined by the scanning velocity |
| $\rho$ | density |
| $\sigma$ | stress tensor |
| $\tau$ | unit vector tangent to the surface, transversal velocity component |
| $\tau$ | dimensionless temperature |
| $\Omega$ | melting heat ratio |
| $\omega$ | dimensionless velocity jump |

*Subscripts*

| | |
|---|---|
| a | ambient |
| L | left |
| l | liquidus |
| M | middle |
| m | melting, solidus |
| NC | no convection |
| R | Rosenthal, right |
| $t$ | time derivative |

*Other*

| | |
|---|---|
| $\nabla$ | nabla operator |
| $\otimes$ | tensor product |
| $\cdot$ | dot product |

# 1. Introduction

The present work focuses on heat and mass transfer at laser processing in the conditions of selective laser melting (SLM) of thin walls. Part I [1] presents the state of the art and the objectives and the experimental approaches and results. The present Part II develops and experimentally validates a two-dimensional computational fluid dynamics (CFD) model of the laser-impact zone. Section 2 explains the theoretical tools used and Section 3 describes the properties of materials accepted for modeling. Section 4 presents and Section 5 discusses the obtained results.

# 2. Theoretical methods

The present work combines numerical CFD modeling to predict laser-material interaction at the given process parameters with simplistic analytical models useful to guide eyes and to solve the inverse problem of finding process parameters to obtain the desirable technological effect. Section 2.1 analyzes the Rosenthal model of the moving heat source. Section 2.2 describes the developed numerical CFD model.

## 2.1. Analytical approaches

Assuming that a point heat source moves along the edge of a thermally thin plate of conductive medium with constant specific heat and thermal conductivity $\lambda$, Rosenthal obtained the well-known analytical solution for two-dimensional temperature field $T$ [2]. This solution is adapted for laser processing of single-track walls as follows [3]:

$$T - T_a = \frac{AP}{\pi\delta\lambda}\exp\left(\frac{Vx}{2a}\right)K_0\left(\frac{Vr}{2a}\right), \quad r = \sqrt{x^2 + z^2}, \tag{1}$$

where $T_a$ is the ambient or preheat temperature, $P$ the incident laser power, $A$ the absorptance, $\delta$ the wall thickness, $V$ the scanning speed, $a$ the thermal diffusivity, and $K_0$ the modified Bessel function of the second kind. Equation (1) is written in frame (OXZ) moving with the laser beam in the negative direction of horizontal axis (OX) and having origin O in the point where the laser beam strikes the plate as shown in Fig. 1. Vertical axis (OZ) is directed downward. The aim is not only calculating temperature field (1) for the given process parameters $P$ and $V$ but also solving the inverse problem of finding the optimal process parameters to obtain the melt pool with the given parameters, length $L$ and depth $D$ shown in Fig. 1. The latter problem is analyzed below.

Increasing SLM productivity requires increasing laser scanning speed $V$, which can attain one meter per second and above. At such a high $V$, the argument of the Bessel function in Eq. (1) becomes much greater than one and the Hankel asymptotic expansion is valid [4]:

$$K_\alpha(x) = \sqrt{\frac{\pi}{2x}}e^{-x}\left(1 + \frac{4\alpha^2 - 1}{8x} + \frac{(4\alpha^2 - 1)(4\alpha^2 - 9)}{128x^2} + O\left(\frac{1}{x^3}\right)\right). \tag{2}$$

Keeping the leading term of expansion (2), Eq. (1) becomes

$$T - T_a = \frac{AP}{\delta\lambda}\sqrt{\frac{a}{\pi V r}}\exp\left(\frac{V(x-r)}{2a}\right). \tag{3}$$

On the positive branch of (OX) axis corresponding to the scanning line behind the heat source, $r = x$ and the exponential term vanishes in Eq. (3). The resulting equation is resolved relative $r$ to obtain the distance between the point source and the intersection of the melting isotherm $T_m$ with the scanning axis behind the source,

$$L_0 = \frac{a}{\pi V}\left(\frac{AP}{\delta\lambda(T_m - T_a)}\right)^2. \tag{4}$$

In the point $(x,z) = (L_0, 0)$, the argument of the Bessel function in Eq. (1) equals

$$\Pi = \frac{VL_0}{2a} = \frac{1}{2\pi}\left(\frac{AP}{\delta\lambda(T_m - T_a)}\right)^2. \tag{5}$$

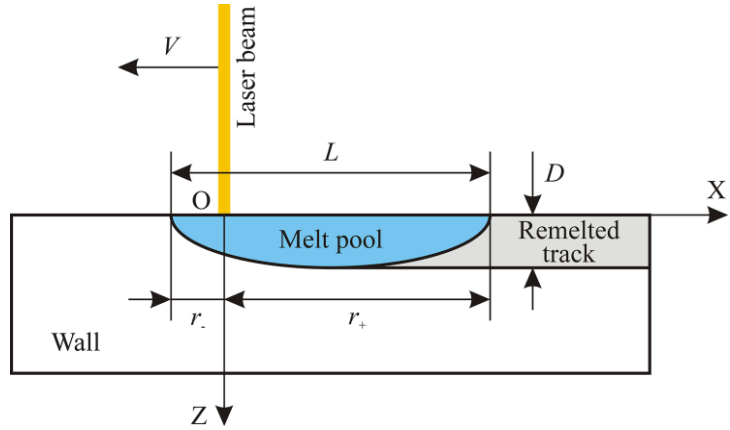

**Fig. 1.** Melt pool in a single-track wall: $L$ is the pool length and $D$ the remelted depth.

According to the above assumptions, Eq. (3) is valid if $\Pi \gg 1$. The value of $L_0$ approximates the length of the melt pool while the value of $\Pi$ is the thermal Peclet number defined by the melt pool length and the scanning velocity. Below, approximation (3) is not employed. However, the obtained complexes (4) and (5) are useful to write rigorous Eq. (1) in the following dimensionless form:

$$\tau = \sqrt{\frac{2\Pi}{\pi}}\exp(\Pi x')K_0(\Pi r'), \tag{6}$$

with dimensionless temperature and coordinates, respectively

$$\tau = \frac{T - T_a}{T_m - T_a}, \quad x' = \frac{x}{L_0}, \quad z' = \frac{z}{L_0}, \quad r' = \frac{r}{L_0}. \tag{7}$$

Equation (6) at $\tau = 1$ defines the melting isotherm. Figure 1 shows the distances $r_+$ and $r_-$ between the point source and the intersection of the melting isotherm with the positive and negative, respectively, branches of axis (OX). These values can be found from condition $x = \pm r$ substituted into Eq. (6), which results in the following equation:

$$\exp(\pm\Pi r')K_0(\Pi r') = \sqrt{\frac{\pi}{2\Pi}}. \tag{8}$$

Equation (6) results in the parametric equation for the melting isotherm,

$$x' = \frac{1}{2\Pi}\ln\frac{\pi}{2\Pi} - \frac{1}{\Pi}\ln(K_0(\Pi r')), \qquad (9)$$

$$z' = \sqrt{r'^2 - x'^2}, \qquad (10)$$

where parameter $r'$ varies from $r_-/L_0$ to $r_+/L_0$.

As shown in Fig. 1, the melt pool length is the sum of the solutions of Eq. (8),

$$L = r_+ + r_-, \qquad (11)$$

Melt pool depth $D$ is defined as the maximum value of coordinate $z$ on the melting isotherm (9)-(10). The necessary condition is a zero of derivative $dz'/dr'$,

$$x' K_1(\Pi r') = r' K_0(\Pi r'), \qquad (12)$$

where $x'$ is defined by Eq. (9). Let solution of Eq. (12) be $r_m$. This value is substituted to Eq. (9) to find $x_m$. Then, the values of $r_m$ and $x_m$ are substituted to Eq. (10) to find $z_m$. Finally, melt pool depth

$$D = z_m. \qquad (13)$$

If $\Pi \gg 1$, one can simplify Eq. (9) by substituting the first two terms of expansion (2) for function $K_0$,

$$x' = r' + \frac{\ln r'}{2\Pi} + \frac{1}{8\Pi^2 r'} + O\left(\frac{1}{\Pi^3}\right). \qquad (14)$$

The first three terms of expansions (2) are necessary for functions $K_0$ and $K_1$ to reduce Eq. (12) to the following equation:

$$r' + r'\ln r' + \frac{3\ln r'}{8\Pi} - \frac{1}{8\Pi} = O\left(\frac{1}{\Pi^2}\right). \qquad (15)$$

One can see that the solution of this equation approaches $r_m/L_0 = e^{-1}$ when $\Pi$ increases. A linear expansion of $\ln r'$ around $r' = e^{-1}$ results that at $\Pi \gg 1$,

$$\frac{r_m}{L_0} = e^{-1} + \frac{1}{2\Pi} + O\left(\frac{1}{\Pi^2}\right). \qquad (16)$$

Substitution of Eq. (16) to Eq. (14) results

$$\frac{x_m}{L_0} = e^{-1} + O\left(\frac{1}{\Pi^2}\right). \qquad (17)$$

Finally, substitution of Eqs. (16) and (17) to Eq. (10) gives the following estimate of the melt depth at $\Pi \gg 1$:

$$\frac{D}{L_0} = \frac{z_m}{L_0} = \frac{1}{\sqrt{e\Pi}} + O\left(\frac{1}{\Pi^{3/2}}\right). \qquad (18)$$

The leading term of this expansion can be written in dimensional form as

$$D_0 = \sqrt{\frac{2}{e\pi}\frac{a}{V}\frac{AP}{\delta\lambda(T_m - T_a)}}. \qquad (19)$$

The value of $D_0$ approximates the melt depth at high thermal Peclet numbers $\Pi$ (see Eq. (5)) like the value of $L_0$ (see Eq. (4)) approximates the melt pool length.

*2.2. CFD model of the melt pool*

A uniform CFD approach is used to model the melt pool and the heat-affected zone (HAZ) without tracing a boundary between them. The solid and liquid phases are distinguished by a temperature threshold. A field of ghost external force is introduced to stop convection in the solid phase. The following partial differential equations describe the conservation laws of mass, momentum, and energy, respectively:

$$\rho_t + \nabla \cdot (\rho \mathbf{u}) = 0, \quad (\rho \mathbf{u})_t + \nabla \cdot \mathbf{\Pi} = 0, \quad E_t + \nabla \cdot \mathbf{Q} = 0, \tag{20}$$

where $\rho$ is the density, $\mathbf{u}$ the flow field, $E$ the energy per unit volume, and index $t$ means a time derivative. It is assumed that flow is laminar. In a viscous conductive medium, the flows of momentum $\mathbf{\Pi}$ and energy $\mathbf{Q}$ are

$$\mathbf{\Pi} = p\mathbf{I} + \rho \mathbf{u} \otimes \mathbf{u} - \boldsymbol{\sigma}, \quad \mathbf{Q} = E\mathbf{u} + \mathbf{\Pi} \cdot \mathbf{u} + \mathbf{q}, \tag{21}$$

where $p$ is the pressure, $\mathbf{I}$ the second-order identity tensor, and $\boldsymbol{\sigma}$ the viscous stress tensor,

$$\boldsymbol{\sigma} = 2\eta \left[ \mathbf{e} - \frac{1}{3} \mathbf{I} \operatorname{tr}(\mathbf{e}) \right], \tag{22}$$

depending on dynamic viscosity $\eta$ and deformation rate tensor $\mathbf{e}$ defined as the symmetric component of tensor $\nabla \mathbf{u}$. The Fourier law gives conductive heat flow,

$$\mathbf{q} = -\lambda \nabla T, \tag{23}$$

where $\lambda$ is the thermal conductivity and $T$ the temperature. Equation of state $p(\rho,T)$ and thermal equation of state $U(\rho,T)$ close the system of Eqs. (20)-(23), where $U = E - \rho u^2/2$ is the internal energy per unit volume.

Assuming low compression and neglecting thermal expansion, the equation of state is defined by the following linear relation between the pressure and the density:

$$p = B\left(\frac{\rho}{\rho_0} - 1\right), \tag{24}$$

where $B$ is the bulk modulus and $\rho_0$ the density at zero pressure. The medium described by Eq. (24) is barotropic because the equation of state is independent of temperature. In such a medium, one can separate elastic $\mathcal{E}$ and thermal $\mathcal{T}$ components of the internal energy,

$$\mathcal{E} = -\int_{\rho_0}^{\rho} p \, d\left(\frac{1}{\rho}\right) = c^2 \left( \ln \frac{\rho}{\rho_0} + \frac{\rho_0}{\rho} - 1 \right), \quad \mathcal{T} = U - \mathcal{E}, \tag{25}$$

where the sound speed squared is

$$c^2 = \frac{dp}{d\rho} = \frac{B}{\rho_0}. \tag{26}$$

The thermal equation of state assumes a constant specific heat $C$ in the solid and liquid states and a uniform release/retain of latent melting enthalpy $L_m$ in the temperature interval from solidus $T_m$ to liquidus $T_l$,

$$T = \begin{cases} \mathcal{T}/C & , \quad \mathcal{T} < CT_m \\ T_m + \dfrac{(T_l - T_m)(\mathcal{T} - CT_m)}{C(T_l - T_m) - L_m} & , \quad CT_m \leq \mathcal{T} \leq CT_l + L_m \\ (\mathcal{T} - L_m)/C & , \quad \mathcal{T} > CT_l + L_m \end{cases}. \tag{27}$$

A laser beam strikes an edge of a thin plate parallel to plane (XZ) in Cartesian coordinates. Figure 2 shows the calculation domain and the coordinate axes. The frame moves with the laser beam scanning the plate with constant velocity $V$ in the direction opposite to axis X. The plate occupies domain $z > 0$. The plate melts around the laser beam forming the melt pool. On the liquid/solid boundary (2 in Fig. 2), the no-slip boundary condition holds meaning equal flow velocities of the solid and liquid phases. On the free melt surface (1 in Fig. 2), the Marangoni condition for shear stress is imposed,

$$\sigma_{n\tau} = \beta \boldsymbol{\tau} \cdot \nabla T, \qquad (28)$$

where $\beta = d\alpha/dT$ is the thermal coefficient of surface tension, $\alpha$ the surface tension, and $\boldsymbol{\tau}$ the unit vector tangent to the surface. It is assumed that the laser beam is Gaussian with nominal radius $r_0$. The heat flow entering the plate through the melt surface 1 is

$$q_z = \frac{AP}{r_0 \delta \sqrt{\pi}} \exp\left(-\frac{x^2}{r_0^2}\right), \qquad (29)$$

where $P$ is the laser power, $A$ the absorptivity, $\delta$ the plate thickness, $z$ the coordinate in the beam direction, and $x$ the coordinate in the scanning direction, see Fig. 2.

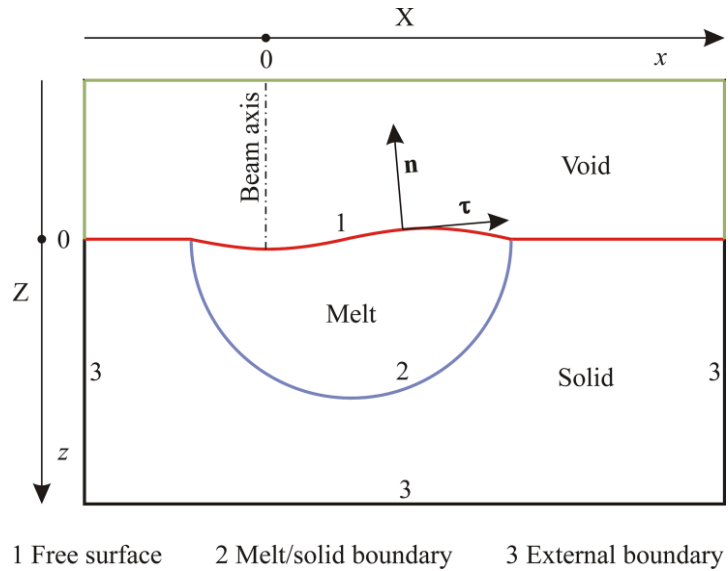

**Fig. 2.** Calculation domain. Boundaries: free surface (1); melting/solidification front (2); external boundaries (3).

Free surface 1 is defined by the volume of fluid (VOF) method with level function $s(x,z)$. The surface corresponds to curve $s = 0$. External normal to the surface $\mathbf{n}$ and unit tangent vector $\boldsymbol{\tau}$ are

$$\mathbf{n} = -\frac{\nabla s}{|\nabla s|}, \quad \boldsymbol{\tau} = \begin{pmatrix} \tau_x \\ \tau_z \end{pmatrix} = \begin{pmatrix} -n_z \\ n_x \end{pmatrix}. \qquad (30)$$

Boundary melt/solid (2 in Fig. 2) is not explicitly traced but is defined by isotherm $T = T_m$. To model solid phase, a ghost external force field is applied in domain $T < T_m$. The ghost field is calculated at every time step to stop convection. This ensures the correct flow field in the solid phase and the no-slip condition on boundary 2.

To reduce the influence of external boundaries (3 in Fig. 2), the so-called non-disturbing boundary conditions are applied [3] derived from the Rosenthal moving point-source solution. These boundary conditions make it possible to minimize the computation domain thus accelerating numerical calculations and increasing the accuracy. On the left and right branches $x$ = const of boundary 3, the non-disturbing boundary condition is [3]

$$-\frac{1}{T-T_a}\frac{\partial T}{\partial x} = \frac{V}{2a}\left(\frac{x}{r}\frac{K_1\left(\frac{Vr}{2a}\right)}{K_0\left(\frac{Vr}{2a}\right)} - 1\right), \tag{31}$$

where $T_a$ is the ambient temperature, $V$ the laser scanning speed, $a = \lambda/C$ the thermal diffusivity, $r^2 = x^2 + z^2$, and $K_0$ и $K_1$ the modified Bessel functions of the second kind. On the bottom branch $z$ = const of boundary 3, the non-disturbing boundary condition is [3]

$$-\frac{1}{T-T_a}\frac{\partial T}{\partial z} = \frac{V}{2a}\frac{z}{r}\frac{K_1\left(\frac{Vr}{2a}\right)}{K_0\left(\frac{Vr}{2a}\right)}. \tag{32}$$

Conservation laws (20) are numerically solved by the finite volume method on a uniform grid with rectangular cells in the calculation domain shown in Fig. 2. The transient calculation scheme results in a steady-state solution. The numerical algorithm is implemented in C++ language and tested [5]. In the finite volume method, the fields of conserved quantities $\rho$, $\rho\mathbf{u}$, and $E$ are specified with the arrays of their average values over the cells.

The fluxes of mass, momentum, and energy through cell boundaries are calculated by a second-order Godunov scheme with a "minmod" approximation of the fluid-dynamic values at the cell boundaries [6] ensuring monotonicity. The "mimnod" approximation results in the values of the fluid-dynamic values at the opposite sides of a cell boundary. The values of the same physical quantity are generally different at the opposite sides. In the Godunov scheme, the resulting "mean" fluid-dynamic parameters on the cell boundary are obtained from the sets of the "left" and "right" values as the solution of the so-called Riemann problem. The Riemann problem for fluid-dynamic equations is an initial-value problem with pricewise constant initial data. It concerns evolution of a jump in initial data. Application of the Riemann problem improves the stability of the numerical scheme and increases the computation accuracy for physical discontinuities [6]. An original Riemann solver is developed for the considered model fluid with equation of state (24) and thermal equation of state (27). Appendix A describes the solver algorithm.

## 3. Material properties

Alloy AlSi10Mg is highly demanded for printing lightweight cellular structures. This work presents examples of single-track thin walls obtained by SLM from this material. Eutectic solder Sn60Pb40 is used for high-speed imaging experiments because foils of the given thickness can be easily fabricated from this soft alloy with a jewelry rolling mill. Pure iron is used for benchmark numerical simulation because it is the base for widely employed steel alloys.

*AlSi10Mg*

Thermophysical properties of alloy AlSi10Mg are taken from Refs. [7-12] and listed in Table 1. Reference density $\rho_0$ corresponds to the room temperature [7]. Specific heat $C$ depends on temperature [9]. The model assumes a constant value of specific heat. The value of specific heat is taken near the solidus point [9], which approximately corresponds to the mean value in the considered temperature interval. Thermal conductivity $\lambda$ is sensitive to the manufacturing method and heat treatment [7]. An anisotropy in thermal conductivity within 10-20% is observed after SLM [7] because of growth of oriented dendrites. In the present work, it is accepted the average between the thermal conductivity values in the building direction and in the direction perpendicular to the building one [10]. Melt viscosity $\eta$ can considerably decrease with temperature [11]. The model assumes a constant viscosity value approximately corresponding to the temperature of 1000 K. The temperature coefficient of surface tension is estimated from the data of Ref. [12].

**Table 1.** Properties of alloy AlSi10Mg accepted for simulation

| Physical quantity | Value | Reference |
|---|---|---|
| Density, $\rho_0$ | 2680 kg/m$^3$ | [7] |
| Solidus temperature, $T_m$ | 830 K | [8] |
| Liquidus temperature, $T_l$ | 869 K | [8] |
| Latent melting heat, $L_m/\rho_0$ | 400 kJ/kg | Al [9] |
| Specific heat, $C/\rho_0$ | 1140 J/(kg K) | Al, 800 K [9] |
| Thermal conductivity, $\lambda$ | 100 W/(m K) | [10] |
| Thermal diffusivity, $a = \lambda/C$ | 32.731 mm$^2$/s | Calculated |
| Dynamic viscosity, $\eta$ | 1.2 mPa s | Al, 1000 K [11] |
| Surface tension, $\alpha$ | 0.85 N/m | Al [12] |
| Thermal coefficient of surface tension, $\beta$ | – 0.2 mN/(m K) | Al [12] |

*Sn60Pb40*

This work accepts the thermophysical properties of Sn60Pb40 alloy reported in Ref. [13]. The absorptance is estimated by the Fresnel equations from the complex refractive index of pure Sn [14]. Assael *et al.* [15] analyzed experimental data on viscosity $\eta$ of this alloy and proposed the following Arrhenius approximation:

$$\log_{10} \frac{\eta}{\eta_0} = -a_1 + \frac{a_2}{T}, \tag{33}$$

where $T$ is the absolute temperature and constants $\eta_0$, $a_1$, and $a_2$ are listed in Table 2. White [16] and Carroll and Warwick [17] reported a complicated temperature dependence of the surface tension with a plateau near the melting point and another plateau above 600 K. In this work, the surface tension is approximated by the following function:

$$\alpha = \frac{\alpha_1 + \alpha_2}{2} - \frac{\alpha_1 - \alpha_2}{2} \frac{T - T_0}{\sqrt{\Delta T^2 + (T - T_0)^2}}, \quad (34)$$

where $\alpha_1$ and $\alpha_2$ are the left and right plateau values, respectively, $T_0$ the inflection temperature, and $\Delta T$ the characteristic width. Figure 3 shows function (34) with the parameters listed in Table 3 and indicates that this function does approximate experimental data [17]. The temperature coefficient of surface tension is given by the derivative of Eq. (1) as follows:

$$\beta = \frac{d\alpha}{dT} = -\frac{\alpha_1 - \alpha_2}{2} \frac{\Delta T^2}{\left(\Delta T^2 + (T - T_0)^2\right)^{3/2}}. \quad (35)$$

The lower full line in Fig. 3 plots Eq. (35). Temperature coefficient $\beta$ is negative and attains the minimum at the inflection temperature.

**Table 2.** Properties of alloy Sn60Pb40 accepted for simulation

| Physical quantity | | Value | Reference |
|---|---|---|---|
| Density, $\rho_0$ | | 8400 kg/m$^3$ | [13] |
| Melting point, $T_m = T_l$ | | 456 K | [13] |
| Latent melting heat, $L_m/\rho_0$ | | 37 kJ/kg | [13] |
| Specific heat, $C/\rho_0$ | | 150 J/(kg K) | [13] |
| Thermal conductivity, $\lambda$ | | 51 W/(K m) | [13] |
| Thermal diffusivity, $a = \lambda/(\rho C)$ | | 40.5 mm$^2$/s | Calculated |
| Absorptance at 1.07 μm, $A$ | | 0.188 | [14], pure Sn |
| Constants to calculate dynamic viscosity $\eta$ by Eq. (33) | $\eta_0$ | 1 mPa s | [15] |
| | $a_1$ | .2266 | |
| | $a_2$ | 280.69 K | |
| Surface tension, $\alpha$ | | Eq. (34) | Present work |
| Thermal coefficient of surface tension, $\beta$ | | Eq. (35) | Present work |

*Iron*

Table 4 lists the properties of iron accepted for modeling. They are selected according to Refs. [9, 18-22]. At elevated temperatures, the density of solid iron drops from approximately 7.8 g/cm$^3$ at 500 K to approximately 7.3 g/cm$^3$ at the melting point 1810 K [18] and becomes around 7.0 g/cm$^3$ in the liquid phase at the melting point [21]. The model nominal density value $\rho_0 = 7.3$ g/cm$^3$ satisfactorily represents both the high-temperature solid and liquid phases. High-temperature heat capacity of iron grows from approximately 35 J/(mol K) at 1200 K to approximately 45 J/(mol K) at the melting point in the solid phase and attains 47 J/(mol K) in the liquid phase [21]. The accepted nominal value of heat capacity shown in Table 4 corresponds to the solid phase at the melting point [21].

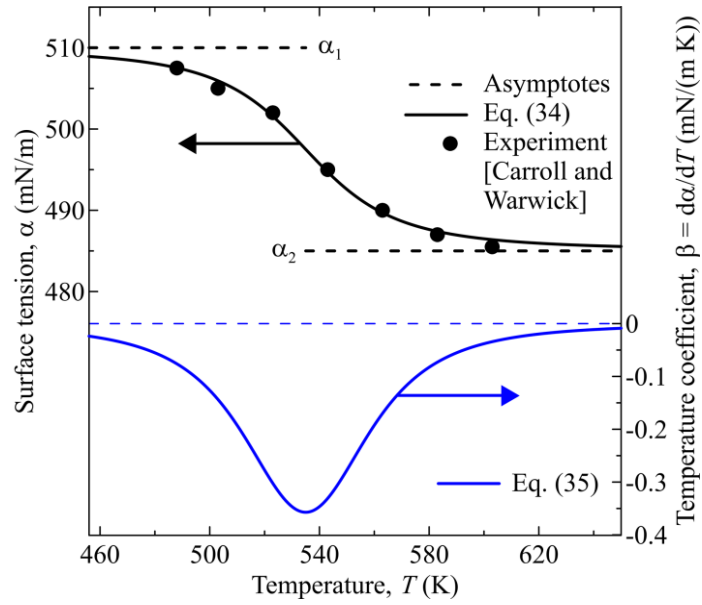

**Fig. 3.** Surface tension $\alpha$ of liquid solder Sn60Pb40 and its temperature coefficient $\beta = d\alpha/dT$: analytical approximations by Eqs. (34) and (35) accepted in the present work (solid lines) compared to experimental data of Carroll and Warwick [17] (points).

**Table 3.** Parameters for Eqs. (34) and (35) approximating the surface tension of liquid solder Sn60Pb40

| Parameter | $\alpha_1$ | $\alpha_2$ | $T_0$ | $\Delta T$ |
|---|---|---|---|---|
| Value | 510 mN/m | 485 mN/m | 535 K | 35 K |

**Table 4.** Properties of pure iron accepted for simulation

| Physical quantity | | Value | Reference |
|---|---|---|---|
| Density, $\rho_0$ | | 7300 kg/m$^3$ | [18] |
| Curie point, $T_c$ | | 1043 K | [19] |
| $\alpha$-$\gamma$ transformation point, $T_{\alpha\gamma}$ | | 1183 K | [19] |
| Melting point, $T_m = T_l$ | | 1810 K | [9] |
| Latent melting heat, $L_m/\rho_0$ | | 247 kJ/kg | [9] |
| Specific heat, $C/\rho_0$ | | 806 J/(kg K) | [20] |
| Thermal conductivity, $\lambda$ | | Eq. (36) | Present approximation of data [19] |
| | | 35 W/(m K) | Accepted constant value |
| Thermal diffusivity, $a = \lambda/(\rho C)$ | | 5.94854 mm$^2$/s | Calculated constant value |
| Constants to calculate dynamic viscosity $\eta$ by Eq. (33) | $\eta_0$ | 1 mPa s | [21] |
| | $a_1$ | .7209 | |
| | $a_2$ | 2694.95 K | |
| Surface tension, $\alpha$ | | 1.8 N/m | [22] |
| Thermal coefficient of surface tension, $\beta$ | | $-0.3$ mN/(m K) | [22] |

Solid-phase thermal conductivity of iron varies between 30 and 40 W/(m K) in the temperature interval from 850 K to the melting point and becomes approximately 40 W/(m K) in the liquid phase at melting [19]. The following equation for thermal conductivity $\lambda$ versus temperature $T$ approximates experimental data [19]:

$$\frac{\lambda}{\text{W/(m K)}} = \begin{cases} 30 + \dfrac{(T-T_c)(T-2800\text{ K})}{38000\text{ K}^2} &, \quad 298\text{ K} < T \leq T_c \\ 30 &, \quad T_c < T \leq T_{\alpha\gamma} \\ 28 - \dfrac{(T-T_{\alpha\gamma})(T-2800\text{ K})}{95000\text{ K}^2} &, \quad T_{\alpha\gamma} < T \leq T_m \\ 40.3 - \dfrac{(T-T_m)(T-3800\text{ K})}{1.66\cdot 10^5\text{ K}^2} &, \quad T_m < T \leq 3100\text{ K} \\ 45.7 - \dfrac{(T-3100\text{ K})(T-2700\text{ K})}{3\cdot 10^5\text{ K}^2} &, \quad 3100\text{ K} < T \leq 6000\text{ K} \end{cases} \quad (36)$$

The value of 35 W/(m K) reported for the solid phase at melting [19] is taken as the nominal value for models with constant thermal conductivity. The viscosity of liquid iron gradually decreases with temperature from 5.443 mPa s at 1850 K to 2.276 mPa s at 2500 K [21]. This function is approximated by Eq. (33) with the parameters listed in Table 4. The surface tension of oxygen-free liquid iron measured in reducing atmosphere slightly decreases with temperature from 1.9 N/m near the melting point to 1.8 N/m around 2100 K [22]. However, the presence of oxygen can reduce surface tension down to approximately 1 N/m near the melting point [22]. In the present work, the surface tension of pure iron is accepted to be 1.8 N/m and its temperature coefficient estimated in the oxygen-free conditions equals to – 0.3 mN/(m K) [22].

One can take into account that according to experimental data [22] the temperature dependence of surface tension may become non-monotonous for iron contaminated with oxygen. In this case, the temperature coefficient may become positive in a temperature interval contiguous to the melting point [22]. Points in Fig. 4 show experimental data [22] on surface tension at oxygen activity of $10^{-11}$. The data at lower and higher temperatures approach two lines (dashed lines). The natural approximation of such data is the lower branch of hyperbola with asymptotes $\alpha = \alpha_1 + \beta_1 T$ and $\alpha = \alpha_2 + \beta_2 T$:

$$\alpha = \frac{\alpha_1 + \alpha_2}{2} + \frac{\beta_1 + \beta_2}{2}T - \frac{1}{2}\sqrt{(\alpha_1 - \alpha_2 + (\beta_1 - \beta_2)T)^2 + 4\alpha_0^2}, \quad (37)$$

where parameter $\alpha_0$ characterizes the distance from the hyperbola to its center. Figure 4 shows hyperbola given by Eq. (37) (upper full line) with the parameters listed in Table 5. One can see that this curve correctly approximates the experimental points. The temperature coefficient of surface tension is given by the derivative of Eq. (37) as follows:

$$\beta = \frac{d\alpha}{dT} = \frac{\beta_1 + \beta_2}{2} - \frac{(\alpha_1 - \alpha_2 + (\beta_1 - \beta_2)T)(\beta_1 - \beta_2)}{2\sqrt{(\alpha_1 - \alpha_2 + (\beta_1 - \beta_2)T)^2 + 4\alpha_0^2}}. \quad (38)$$

The lower full line in Fig. 4 plots Eq. (38). Temperature coefficient $\beta$ is positive near the melting point 1810 K, decreases with temperature and becomes negative above approximately 2000 K.

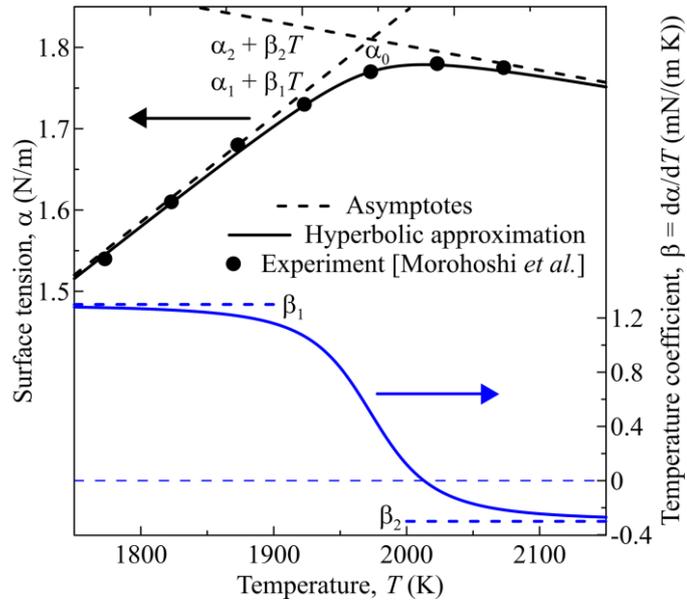

**Fig. 4.** Hyperbolic approximation for surface tension of liquid iron $\alpha$ and its temperature coefficient $\beta = d\alpha/dT$ (solid lines) compared to experimental data of Morohoshi *et al.* [22] (points) at oxygen activity of $10^{-11}$.

**Table 5.** Parameters for the hyperbolic approximation of surface tension

| Parameter | $\alpha_0$ | $\alpha_1$ | $\alpha_2$ | $\beta_1$ | $\beta_2$ |
|---|---|---|---|---|---|
| Value | 0.04 | -0.7549 | 2.4019 | 1.3 | -0.3 |
| Unit | N/m | | | mN/(m K) | |

## 4. Results

### 4.1. Rosenthal model

According to Eq. (6), the dimensionless temperature field of a moving point source depends on the unique parameter $\Pi$ given by Eq. (5). Parameter $\Pi$ contains laser processing parameters and material properties. When $\Pi \gg 1$, this parameter is essentially the thermal Peclet number defined by the melt pool length and the laser scanning velocity. Numerical calculations are accomplished in the range $1/32 \leq \Pi \leq 128$. Melt pool length $L$ and depth $D$, distances from the point source to the pool front $r_-$ and back $r_+$, $x$-coordinate of the maximum pool depth $x_m$, and aspect ratio $L/D$ are listed in Appendix B and plotted versus $\Pi$ in Fig. 5. One can see that pool length $L$ tends to $L_0$ and melt depth $D$ to $D_0$ as $\Pi$ increases. Distance $r_-$ approaches zero while distance $r_+$ approaches $L$. The numerical data and the plots indicate that in the range $\Pi \geq 1/2$, Eq. (4) approximates the melt pool length with the relative error within 5%. In the range $\Pi \geq 4$, Eq. (19) approximates the melt depth with the relative error within 10%.

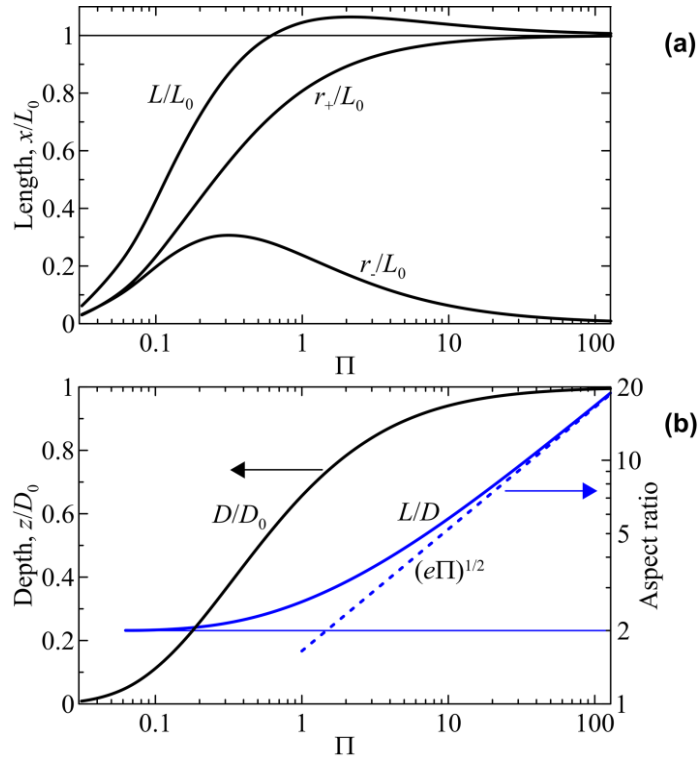

**Fig. 5.** Rosenthal model: Dimensionless melt pool length (a) and depth (b) and aspect ratio (b) versus thermal Peclet number Π.

Figure 6 shows the calculated profiles of the melt pool. At Π < 1, the melt pool is approximately hemi-circular. The profile becomes more elongated with increasing Π. The latter tendency is illustrated by aspect ratio *L/D* plotted in Fig. 5b. At Π >> 1, the aspect ratio becomes

$$\frac{L_0}{D_0} = \sqrt{e\Pi} .  \tag{39}$$

This function is shown by dashed line in Fig. 5b indicating that the aspect ratio increases as the square root of the thermal Peclet number. According to Eq. (5), the thermal Peclet number Π is independent of scanning velocity *V* and proportional to laser power *P* squared. Equation (19) indicates that at high Π, melt depth *D* becomes proportional to the linear energy density

$$\text{LED} = \frac{P}{V} .  \tag{40}$$

Thus, LED is an important parameter essentially influencing laser processing of thin walls.

*4.2. CFD simulation*

In computational fluid dynamic (CFD) simulation, laser processing conditions are specified by scanning speed *V* and absorbed specific power $AP/\delta$. These parameters correspond to experiments with AlSi10Mg and Sn60Pb40. CFD simulation is accomplished for the studied processing regime for AlSi10Mg and four studied regimes for Sn60Pb40. In addition, laser processing of iron is simulated with a set of parameters typical for SLM. The studied sets of parameters *V* and $AP/\delta$ are listed in Tables 6 and 7. Table 6 reports more details on the calculation scheme for selected cases including the size of the calculation domain, grid size, and time *t* to obtain a steady-state numerical solution. To understand the influence of convection in the melt pool, the same CDF computations are repeated in the conditions of intentionally stopped convection. In the simplified model of no convection, temperature coefficient of surface tension $\beta$ is set to 0. This eliminates the Marangoni effect being the driving force of convection. To understand the influence of the latent heat of melting, the calculations are repeated for the constant values of specific heat *C* and thermal conductivity $\lambda$ and zero value of latent melting heat $L_m$. The latter model is essentially the Rosenthal model.

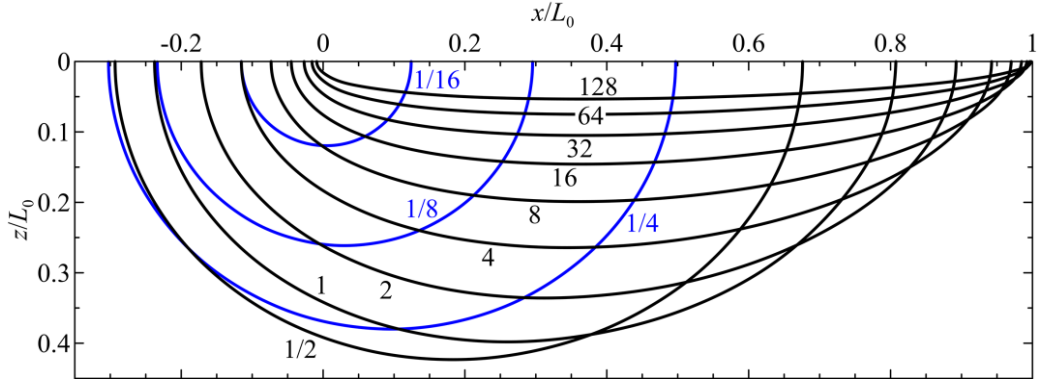

**Fig. 6.** Rosenthal model: Dimensionless melt pool profiles corresponding to the indicated values of thermal Peclet number $\Pi$.

**Table 6.** Parameters for selected CFD calculations

| Material | AlSi10Mg | Sn60Pb40 | Fe | Fe + O |
|---|---|---|---|---|
| Absorbed specific power, $AP/\delta$ | 1 MW/m | 85.7 kW/m | 0.5 MW/m | |
| Scanning speed, *V* | 1.5 m/s | 0.5 m/s | | |
| Calculation domain | 5.12 mm × 320 μm | 10.24 mm × 640 μm | 640 μm × 128 μm | |
| Grid size | 2048 × 128 | 2048 × 128 | 1280 × 256 | |
| Steady-state time, *t* | 10 ms | 30 ms | 2 ms | |
| Calculated melt length, *L* | 3.9575 ± 0.003 mm | 5.86 ± 0.005 mm | 442.8 ± 0.5 μm | 434.3 ± 0.5 μm |
| Calculated melt depth, *D* | 130 ± 3 μm | 220 ± 5 μm | 41.5 ± 0.5 μm | |

Figures 7-9 show the calculated fluid-dynamic fields for the selected cases listed in Table 6. Parts (a) of these figures show the temperature distribution in the computation domain. A thicker line draws the solidus isotherm bounding the melt pool. Parts (b)-(e) show distributions in the melt pool. Streamlines are drawn both in the frame moving with the laser beam (c) and in the laboratory frame (d). Parts (e) show the absolute value of flow velocity. Figures 7d and 8d clearly indicate two vortices formed by the Marangoni force applied to the free surface of the melt pool in the direction from the laser beam to the pool boundary. In Figs. 7c and 8c, the vortices are manifested by deviation of the stream lines from straight translation lines. This means that the flow velocity of circular motion in the vortices is essentially lower than the laser scanning velocity $V$. Despite the tendency of circular motion in the vortices, a point of the melt hardly completes a revolution.

**Table 7.** Melt depth $D$ calculated by the full CFD model, the CFD model with intentionally stopped convection and the Rosenthal model

| Material | Absorbed specific power, $AP/\delta$ | Scanning speed, $V$ | Calculated melt depth | | | $D_{NC}/D_R$ |
|---|---|---|---|---|---|---|
| | | | Full CFD model, $D$ | No convection, $D_{NC}$ | Rosenthal, $D_R$ | |
| Fe | 0.5 MW/m | 0.5 m/s | 41.5 ± 0.5 µm | 46 ± 1 µm | 51 | 0.90 ± 0.02 |
| AlSi10Mg | 1 MW/m | 1.5 m/s | 130 ± 3 µm | 150 ± 3 µm | 195 | 0.77 ± 0.02 |
| Sn60Pb40 | 27.6 kW/m | 0.2 m/s | 144 ± 2 µm | 144 ± 2 µm | 256 | 0.56 ± 0.01 |
| | 55.3 kW/m | | 340 ± 10 µm | 340 ± 10 µm | 620 | 0.55 ± 0.02 |
| | 85.7 kW/m | | 550 ± 20 µm | 550 ± 20 µm | 1005 | 0.55 ± 0.02 |
| | | 0.5 m/s | 220 ± 5 µm | 220 ± 5 µm | 400 | 0.55 ± 0.01 |

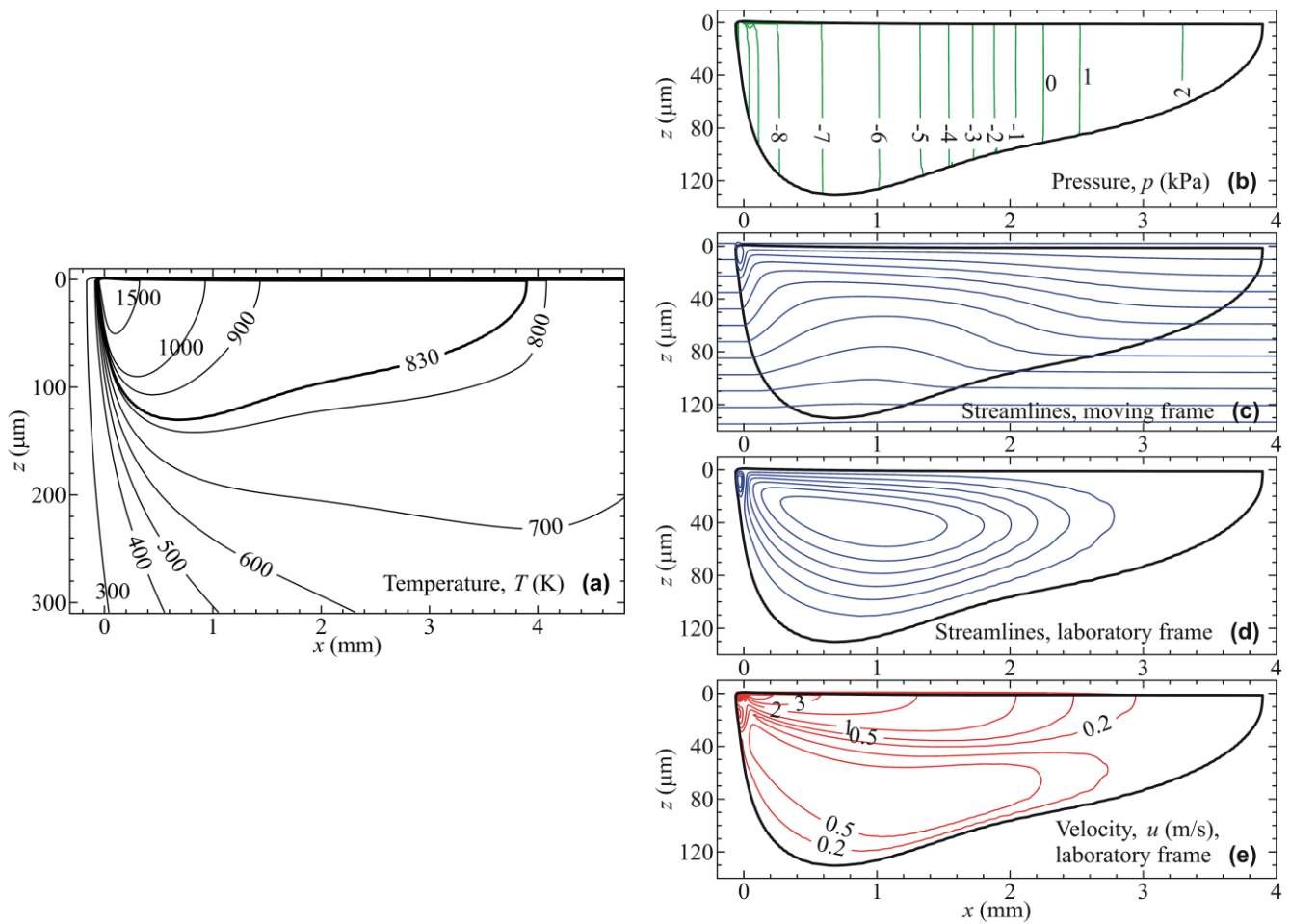

**Fig. 7.** Numerical CFD simulation of the laser-impact zone in AlSi10Mg at absorbed specific laser power $AP/\delta$ = 1 MW/m and scanning speed $V$ = 1.5 m/s: temperature distribution in the melt pool and heat-affected zone (a) and fluid flow parameters in the melt pool (b)-(e).

Figure 9 shows the influence of a surface-active impurity on the melt pool. The addition of oxygen to iron changes the temperature dependence of surface tension as indicated in Fig. 4. At oxygen activity of $10^{-11}$, the temperature coefficient of surface tension becomes positive at temperatures below approximately 2000 K while it is still negative at higher temperatures. The CFD simulation indicates that the melt flow in pure iron (left column in Fig. 9) is qualitatively similar to the flows shown in Figs. 7 and 8 while there are two additional vortices in the presence of oxygen, see the right of Fig. 9d. The additional vortices arise in the periphery of the melt pool. The rotation direction in the additional low-temperature vortices is opposite to that in the principal high-temperature vortices. As a result, a descending melt flow is formed in the melt pool behind the laser spot centered at $x = 0$. This flow results in the additional depression in the melting isotherm clearly visible in the right column of Fig. 9.

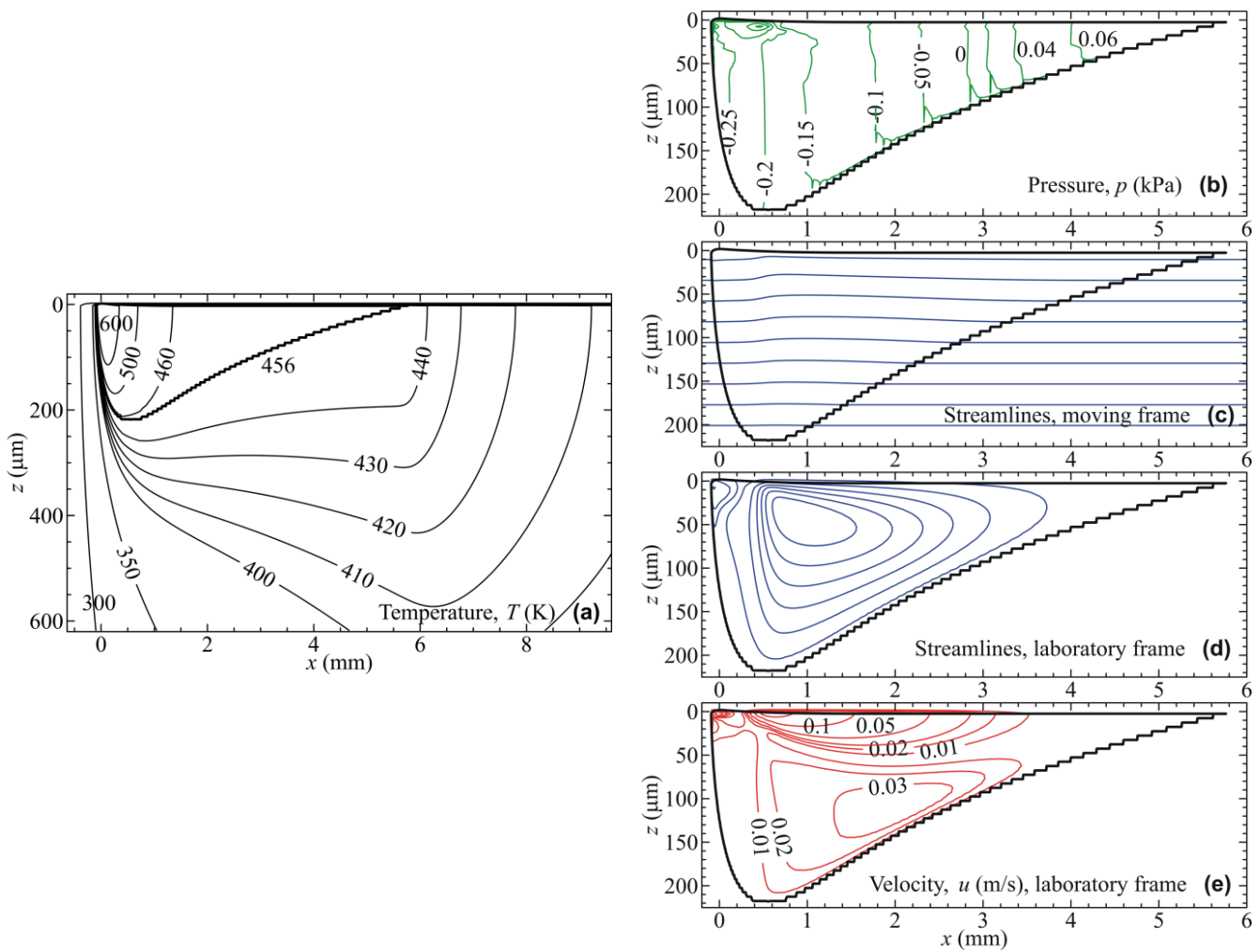

**Fig. 8.** Numerical CFD simulation of the laser-impact zone in Sn60Pb40 at absorbed specific laser power $AP/\delta$ = 85.7 kW/m and scanning speed $V = 0.5$ m/s: temperature distribution in the melt pool and heat-affected zone (a) and fluid flow parameters in the melt pool (b)-(e).

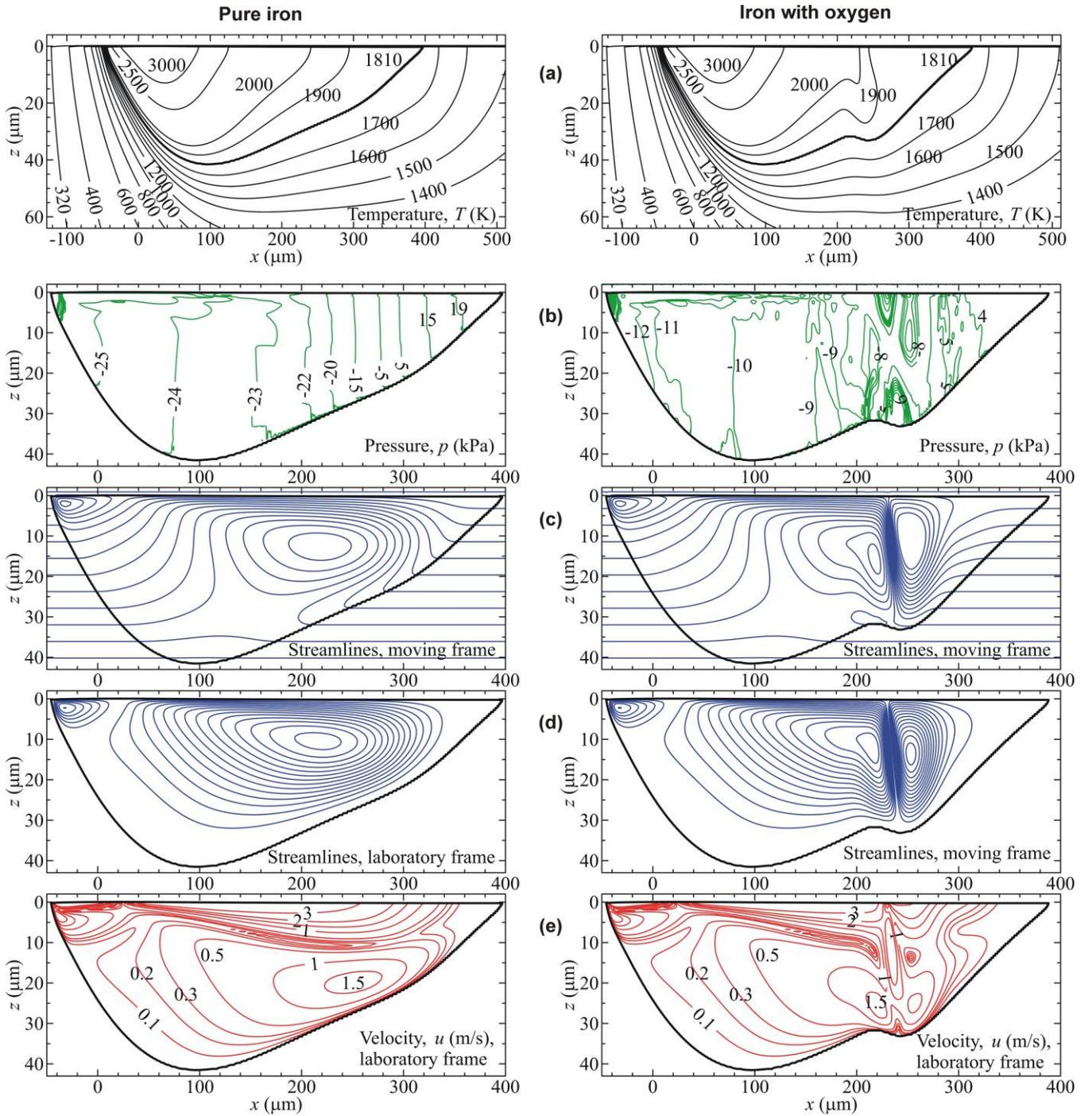

**Fig. 9.** Influence of surface-active oxygen impurity on the melt pool in iron: pure iron (left) and iron with $10^{-11}$ oxygen activity (right). Numerical CFD simulation of the laser-impact zone at absorbed specific laser power $AP/\delta = 0.5$ MW/m and scanning speed $V = 0.5$ m/s: temperature distribution in the melt pool and heat-affected zone (a) and fluid flow parameters in the melt pool (b)-(e).

## 5. Discussion

*5.1. Comparison of the CFD model with experiments*

Figure 10 shows the Sn60Pb40 surface around the laser spot [1]. The dark domain is likely to be the free surface of the melt pool. The surrounding solid surface is rough. Therefore, there are numerous bright dots. This experimental image is compared with the solidus isotherm (red line) calculated with the CFD model at the same scanning speed $V = 0.2$ m/s and absorbed specific power $AP/\delta = 27.6$ kW/m. The calculated contour of the melt pool is similar to the experimentally observed one but the calculated size is approximately 20% lower. The reason for this discrepancy can be underestimation of the accepted absorptivity value $A$ of Sn60Pb40 melt, see Table. 2. The value is taken for pure Sn at the room temperature. It is probable that the addition of Pb and the rise of temperature significantly increase the absorptivity.

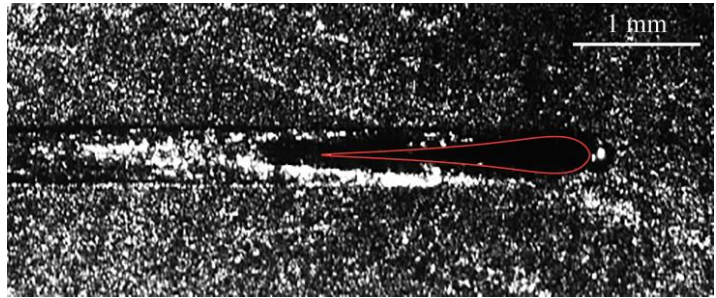

**Fig. 10.** Calculated solidus isotherm (red line) superposed on a video frame of Sn60Pb40 at the scanning speed $V = 0.2$ m/s and laser power $P = 50$ W.

Reference [1] visualizes boundaries between layers in single-track walls of AlSi10Mg and indicates that the depth of the melt pool is approximately between 150 and 180 μm. The calculated melt pool depth is around 130 μm, see Table 7. Thus, the CFD model underestimates the melt pool depth in AlSi10Mg too. This can be explained by underestimation of the absorptivity. Besides, a depression deformation of the melt pool surface around the laser spot was observed [5]. The depression can form due to the recoil pressure of vapor and result in deepening the melt pool.

Figure 11 compares the experimental and calculated melt pool sizes for all the four studied laser processing regimes of Sn60Pb40. The experimental geometry of laser processing corresponds to the right sketch in Fig. 12 while the CFD simulation geometry is shown by the left sketch in the same figure. Therefore, the laser power $P$ accepted for the simulation is a half of the experimental laser power and the calculated melt width in Fig. 11a is taken as $W = 2D$, where $D$ is the melt depth listed in Table 6. The results are plotted versus the linear energy density LED calculated by Eq. (40) because this parameter is commonly considered as the principal characteristics of laser processing intensity [23,24]. Figure 11 shows the same tendencies for the experimental data and CFD simulation. The difference between the experiments and the calculations can attain approximately 20%. The CFD model systematically underestimates both the melt pool width and length in agreement with Fig. 10. Such a disagreement can be reduced by taking a more realistic laser absorptivity value in the model. Therefore, one can conclude that the developed CFD model is experimentally validated.

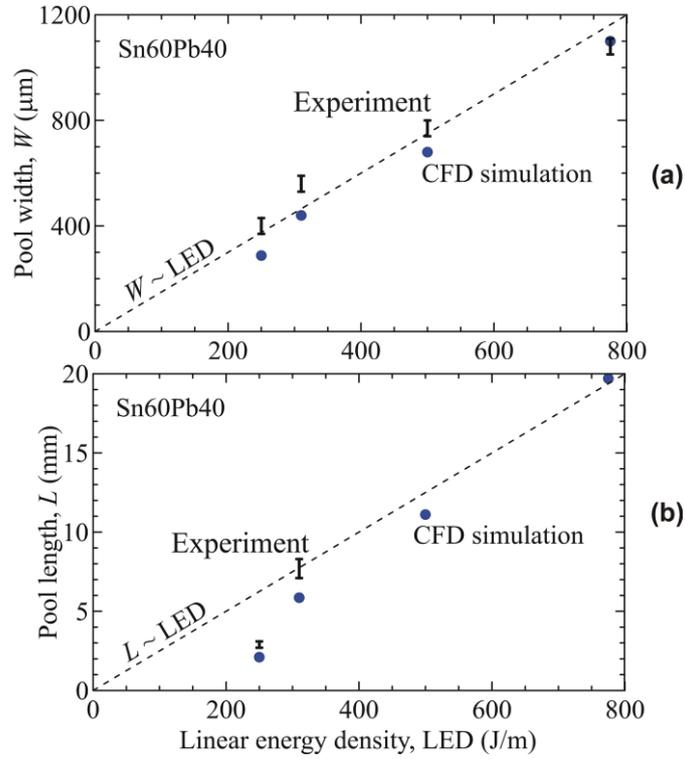

**Fig. 11.** Comparison between experimental (vertical bars) and calculated (circles) melt pool width $W$ (a) and length $L$ (b). The dashed line traces proportionally to linear energy density LED.

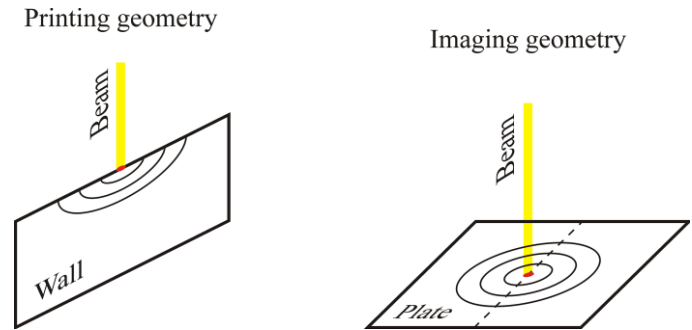

**Fig. 12.** Modeling and SLM printing geometry (left) versus high-speed imaging geometry [1] (right).

*5.2. Linear energy density*

Figure 11a shows the proportionality of the both experimental and calculated melt width $W$ to linear energy density LED. This means that the melt depth $D$ is proportional to LED in the geometry of SLM shown in the left sketch of Fig. 12. The Rosenthal model also indicates the proportionality between the melt depth and LED, see Eq. (19). Figure 11b shows that melt length increases with LED but there is no proportionality between these values. This observation agrees with the Rosenthal model. Indeed, Eq. (4) indicates that melt length $L$ is proportional to $P^2/V = \text{LED} \times P$. The estimation of melt depth is important in SLM to choose the layer thickness and avoid lack-of-fusion defects. The results of the present study indicate that the melt depth is proportional to LED for a given material. This can be useful when optimizing the SLM process.

*5.3. Convection in the melt pool*

The present work theoretically estimates Marangoni convection in the melt pool when obtaining single-track walls by SLM. The thermocapillary force is applied to the melt free surface in the direction of the surface tension gradient, see Eq. (28). In the typical case where surface tension decreases with temperature, the thermocapillary force results in formation of two outward vortices in the melt pool as shown in Figs. 7d and 8d and Fig. 9d on the left. In the outward vortex, melt flows from the laser spot to the melt pool periphery near the free surface and returns near the pool bottom. Surface-active impurities can make the surface tension-temperature function non-monotonous giving raise additional vortices with the opposite inward flow direction as shown on the right in Fig. 9d. Kovalev and Gurin [25] also predicted multi-vortex convection in a laser melt pool. However, their multi-vortex convection was not related to the surface tension-temperature non-monotony.

The intensity of convection is commonly characterized by the Reynolds number estimating the ratio of the dynamic pressure to viscous pressure,

$$\text{Re} = \frac{\rho_0 u D}{\eta}, \qquad (41)$$

with the characteristic values of density $\rho_0$, flow velocity $u$, pool size $D$, and dynamic viscosity $\eta$. Table 8 estimates the Reynolds number for selected cases. Here, the characteristic pool size $D$ is the melt depth, the characteristic flow velocity $u$ is the local maximum of velocity absolute value at the bottom of the vortex behind the laser spot, and the values of melt properties, $\rho_0$ and $\eta$, are taken around the solidus temperature. Table 8 shows that all the obtained Re values are much greater than one indicating developed melt flow. On the other hand, the values are well below 1000. Therefore, arising a turbulence in the melt pool is hardly possible. This validates the model assumption of laminar flow.

**Table 8.** Estimation of the Reynolds and thermal Peclet numbers in the melt-pool flow

| Material | AlSi10Mg | Sn60Pb40 | Fe |
| --- | --- | --- | --- |
| Absorbed specific power, $AP/\delta$ | 1 MW/m | 85.7 kW/m | 0.5 MW/m |
| Scanning speed, $V$ | 1.5 m/s | 0.5 m/s | |
| Melt depth, $D$ | 130 μm | 220 μm | 41.5 μm |
| Flow velocity, $u$ | 0.83 m/s | 0.037 m/s | 1.66 m/s |
| Dynamic viscosity, $\eta$ | 1.2 mPa s | 2.45 mPa s | 5.86 mPa s |
| Thermal diffusivity, $a$ | 32.731 mm$^2$/s | 40.5 mm$^2$/s | 5.94854 mm$^2$/s |
| Reynolds number, Re | 240 | 28 | 85 |
| Thermal Peclet number, Pe | 3.3 | 0.2 | 12 |

Another important dimensionless characteristics of the flow in the melt pool is the thermal Peclet number

$$\text{Pe} = \frac{uD}{a}, \tag{42}$$

where $a$ is the thermal diffusivity. The value of Pe estimates the ratio of convective heat transfer to conductive one. The selected values of Pe are listed in Table 8. They are obtained from the above-mentioned values of $D$ and $u$ and the characteristic values of thermal diffusivity taken around the solidus temperature. The value of Pe is below one for alloy Sn60Pb40. The low value of Pe is the result of a low temperature coefficient of surface tension, low temperature gradients, and high thermal diffusivity. One can suppose that the convective flow is not important for melt pool formation in this alloy. The values of Pe are above one for alloy AlSi10Mg and iron. Thus, convective flow can significantly change heat transfer in the melt pool.

To estimate the influence of convection, Fig. 13 compares the melt pool profiles calculated by the full CFD model (thick lines) with the profiles calculated by CFD model with intentionally stopped convection (blue lines). Figures 13c-13f show that convection does not influence the melt pool profile in Sn60Pb40. This result is expected assuming the low value of Pe. However, Figs. 13a and 13b show that even in AlSi10Mg and Fe the influence of convection on the melt pool size is not as high as would be expected from the values of Pe listed in Table 8. When convection is intentionally stopped, the melt depth and length change within 10%.

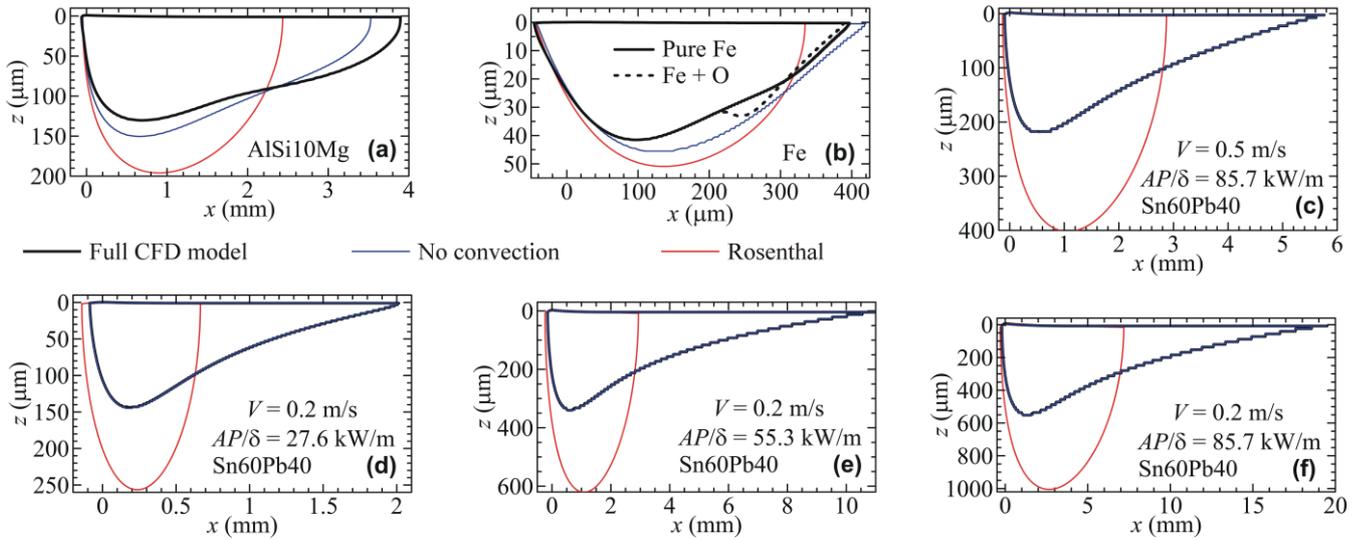

**Fig. 13.** Solidus isotherms calculated according to the following models: full CFD model (thick line), CFD model with intentionally stopped convection (blue line), and Rosenthal model (red line). The material and laser processing parameters are indicated in plots (a)-(f).

In a vortex, the direction of melt flow near the free surface is opposite to that near the melt pool bottom. Therefore, the two convective heat fluxes partly eliminate each other as shown in Fig. 14. Conductive heat flux is always directed from the laser spot to the melt pool periphery. The partial elimination of the four convective heat fluxes shown in Fig. 14 results in the fact that the influence of convection on the temperature field and the melt pool size is low in the considered conditions of SLM. Such conditions are likely obtained due to a low depth of the melt pool of the order of 100 μm. At such a low size, viscous forces should considerably slow the melt flow. For example, it is known that convective heat transfer dominates in melt pools with the depth of the order of 1 mm formed in laser cladding [25].

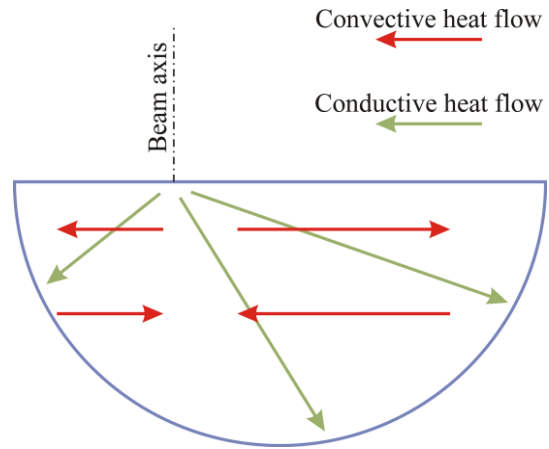

**Fig. 14.** Heat transfer in the melt pool with the pairs of opposite convective heat fluxes.

In summary, the use of a simplified model of heat transfer in the laser-impact zone is acceptable in the conditions of SLM to calculate the shape and size of the melt pool. This model assumes conductive heat transfer and neglects convective heat transfer. In the considered examples, these assumptions assure the relative error within 10% in estimating the melt pool depth and length. Such an accuracy can be sufficient for technologists when optimizing SLM process parameters.

Melt-pool convection assures not only heat transfer but also mass transfer. In SLM, the mass transfer can be important when using powder mixtures of different components for in-situ alloying or to obtain metal matrix composites. In-situ alloying of powder blends is widely applied in coaxial laser cladding and there are successful examples of such a technique in SLM [26]. The examples listed in Table 8 indicate that the flow velocity in the melt pool is around or less than the laser scanning speed. The consequence is that the most of streamlines do not form a loop in Figs. 7c, 8c, and 9c. This means insufficient mixing in the melt pool in SLM. One cannot expect a complete homogenization of chemical composition when using powder blends. On the other hand, such conditions can be favorable for obtaining heterogeneous structures including metal matrix composites.

*5.4. Influence of the latent heat of melting*

Figure 13 shows that the Rosenthal model considerably overestimates melt depth $D$ and underestimates melt length $L$. The only difference between the Rosenthal model and the satisfactory CFD model of conductive heat transfer is neglecting the latent heat of melting. Thus, one can conclude that this material property is important in estimating the melt pool size. The contribution of the latent melting heat $L_m$ can be estimated by the ratio of $L_m$ to the enthalpy of heating up to the melting point. The melting heat ratio is defined as

$$\Omega = \frac{L_m}{C(T_m - T_a)}. \tag{43}$$

Table 9 lists this value for the materials studied in the present work. There is an evident correlation between the value of Ω and the difference between the full CFD model and the Rosenthal model in Fig. 13. The maximum difference is observed for Sn60Pb40 with the maximum value of Ω while the minimum difference is observed for Fe with the minimum value of Ω.

**Table 9.** Melting heat ratio

| Material | AlSi10Mg | Sn60Pb40 | Fe |
|---|---|---|---|
| Melting heat ratio, Ω | 0.6595 | 1.5611 | 0.2027 |

The analytical Rosenthal model would be very useful for estimating the melt pool sizes if it did not result in unsatisfactory error because of neglecting the latent melting heat. However, the revealed correlation between the error of the Rosenthal model and parameter Ω can be explored to develop a satisfactory analytical tool. Therefore, this correlation worthens a deeper study. Below, we focus on melt depth $D$. To exclude numerous parameters related to melt flow, Figure 15 compares the Rosenthal model with the CFD model with intentionally stopped convection. The values of $D_{NC}$ and $D_R$ are taken from Table 7. Figure 15 shows that the ratio of these values perfectly correlates with the melting heat ratio for all the cases of SLM studied in the present work.

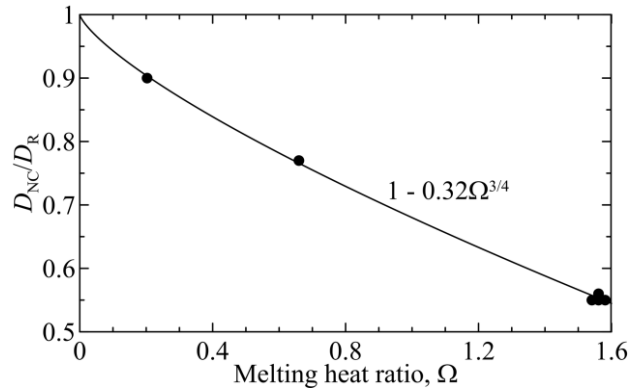

**Fig. 15.** Correction factor $D_{NC}/D_R$ for estimating the melt depth with the Rosenthal model.

*5.5. Analytical model for estimating the melt depth*

Figure 15 proposes the following expression for the melt depth:

$$D = (1 - 0.32\Omega^{3/4})D_R, \tag{44}$$

where $D_R$ is the melt depth calculated by the Rosenthal model. The first approximation to $D_R$ can be the value of $D_0$ given by Eq. (19). The rigorous approach includes calculating the value of Π by Eq. (5) and estimating the value of $D_R$ with the plot of Fig. 5b. Table 10 demonstrates the sequence of calculations applied to the SLM cases studied in the present work. Figure 16 compares the obtained values of melt depth (blue bars), the corresponding values numerically calculated with the CFD model (black bars), and the experimentally measured melt depth (red bars). It is accepted that melt depth $D$ in the SLM geometry shown on the left of Fig. 12 is equal to the half of melt width $W/2$ in the present experiment geometry shown on the right of Fig. 12 while laser power $P$ in the SLM geometry is the half of the experimental power shown near the red bars in Fig. 16. The values of $W$ are taken from Ref. [1].

**Table 10.** Estimation of the melt depth with the analytical model.

| Material | $AP/\delta$ | $V$ (m/s) | $AP/(V\delta)$ (MJ/m$^2$) | $\Pi$ Eq. (5) | $D_0$ (μm) Eq. (19) | $D_R$ (μm) Fig. 5b | $D$ (μm) Eq. (44) |
|---|---|---|---|---|---|---|---|
| Fe | 0.5 MW/m | 0.5 | 1 | 14.21 | 54.40 | 52 | 47 |
| AlSi10Mg | 1 MW/m | 1.5 | 0.67 | 56.23 | 198.5 | 196 | 150 |
| Sn60Pb40 | 27.6 kW/m | 0.2 | 0.138 | 1.867 | 335.7 | 255 | 141 |
| | 55.3 kW/m | | 0.277 | 7.496 | 672.5 | 619 | 342 |
| | | | 0.429 | 18.00 | 1042 | 1010 | 559 |
| | 85.7 kW/m | 0.5 | 0.171 | 18.00 | 416.9 | 404 | 223 |

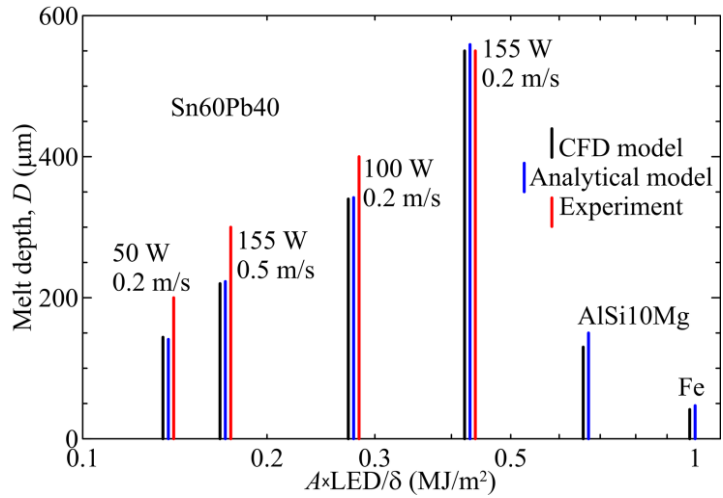

**Fig. 16.** Comparison of the CFD model (black bars), analytical model (blue bars), and experiments [1] (red bars) for the melt depth $D$. Materials and experimental values of laser energy $P$ and scanning velocity $V$ are indicated near the corresponding bars. The horizontal axis shows the value of absorbed linear energy density per unit wall thickness accepted in the models.

Figure 16 indicates that the developed analytical model satisfactorily agrees with the CFD model. The relative difference between the models is within 10% in the studied laser processing conditions typical for SLM. The models correctly follow the experimental results. A systematic deviation of the models from the experiments may arise because of the error in estimating laser absorptivity $A$. Figure 16 demonstrates that the melt depth is approximately proportional to the linear energy density LED for the given material while one cannot compare different materials based on the value of LED.

## 6. Conclusion

- An original CFD model is developed assuming two-dimensional heat conduction and laminar viscous flow of incompressible fluid in a thin wall. The Marangoni thermocapillary effect is the driving force for convection.
- The obtained CFD model equations are numerically solved in a frame moving with the laser beam to obtain steady-state distribution of fluid-dynamic parameters. A low artificial compressibility is introduced to facilitate calculation of the pressure field. The transient conservation laws for mass, momentum, and energy are numerically solved by a second-order Godunov finite-volume method using an original Riemann solver developed for the applied equation of state.
- The CFD model is validated by comparison with experiments [1]. The model systematically underestimates both the melt pool width and length. The difference can attain approximately 20%. Such a disagreement can be reduced by taking a more realistic laser absorptivity value in the model.
- CFD calculations show that if surface tension decreases with temperature, the thermocapillary force results in formation of two outward vortices in the melt pool. Surface-active impurities can make the surface tension-temperature function non-monotonous giving raise additional vortices with the opposite inward flow direction.
- To understand the influence of convection in the melt pool, a reduced CFD model is tested where convection is intentionally stopped. In the considered SLM laser processing regimes, neglecting convection results in the relative error within 10% in estimating the melt pool depth and length. Thus, influence of melt convection on the melt pool sizes is not considerable in the conditions of SLM.
- Numerical CFD simulation indicates that the flow velocity in the melt pool is around or less than the laser scanning speed. This means insufficient mixing in the melt pool. One cannot expect a complete homogenization of chemical composition when using powder blends in SLM. On the other hand, such conditions can be favorable for obtaining heterogeneous structures including metal matrix composites.
- Comparison of the experimental and CFD results with the Rosenthal analytical model indicates that the latter can considerably underestimate melt pool length and overestimate the depth. The reason is neglecting the latent heat of fusion in the Rosenthal model. Thus, the latent heat of fusion appears to be an important parameter in the studied conditions.
- The correlation between the error of the Rosenthal model and the latent heat of fusion is carefully studied resulting an analytical model for estimating the melt pool depth. In the studied SLM cases, the accuracy of the developed analytical model is within 10% relative the full CFD model.

- The obtained experimental [1] and theoretical results indicate that the melt depth is approximately proportional to linear energy density LED in the typical conditions of SLM. This can be useful when optimizing the SLM process.

**CRediT authorship contribution statement**

**Andrey V. Gusarov:** Writing – original draft, Methodology, Software, Investigation. **Tatiana V. Tarasova:** Writing – review & editing, Validation. **Sergey N. Grigoriev:** Writing – review & editing, Conceptualization.

**Declaration of competing interest**

The authors declare that they have no known competing financial interests or personal relationships that could have appeared to influence the work reported in this paper.

**Data availability**

Data will be made available on request.

**Acknowledgments**

This work was supported by the Russian Science Foundation (Grant Agreement No. 23-19-00334 dated 05/15/2023, https://rscf.ru/project/23-19-00334/).

**Appendix A: Riemann solver**

*A1. Riemann problem*

The Riemann problem is a one-dimensional initial-value problem for compressible fluid flow. The initial distributions of fluid parameters are pricewise-constant with the constant values separated by a jump. The principal interest is the splitting of the initial jump. The solution of the Riemann problem is widely employed in numerical methods for fluid dynamics. The aim is to find fluid parameters around the initial jump. An arbitrary jump generally splits into three jumps. One of them is the so-called contact jump and the other two can be shocks or rarefaction fans [27]. Section A2 considers general laws for shocks and rarefaction fans and derives specific relations for the used equation of state. Section A3 describes the developed algorithm for numerical solution. The algorithm is implemented as a Riemann solver written in C++. The present CFD simulation uses the Riemann solver.

*A2. Jump conditions at the shocks and rarefaction fans*

The so-called Rankine-Hugoniot conditions impose the following constraints for the fluid-parameter jumps at the shock front [27]:

$$\rho_1 u_1 = \rho_2 u_2, \tag{A1}$$

$$\rho_1 u_1^2 + p_1 = \rho_2 u_2^2 + p_2, \tag{A2}$$

$$\frac{H_1}{\rho_1} + \frac{u_1^2}{2} = \frac{H_2}{\rho_2} + \frac{u_2^2}{2}, \tag{A3}$$

where enthalpy

$$H = U + p. \tag{A4}$$

Here, the flow velocity is measured relative the moving shock front. Index 1 corresponds to the state before the shock while index 2 designates the values behind the shock. Elimination of $u_1$ и $u_2$ from Eqs. (A1)-(A3) results in the so-called Hugoniot adiabat [27]

$$\frac{H_2}{\rho_2} - \frac{H_1}{\rho_1} = \frac{1}{2}\left(\frac{1}{\rho_1} + \frac{1}{\rho_2}\right)(p_2 - p_1). \tag{A5}$$

The Riemann invariants are defined as [27]:

$$J_\pm = u \pm \int \frac{c\mathrm{d}\rho}{\rho} = u \pm c \ln \rho. \tag{A6}$$

Consider a one-dimensional rarefaction fan propagating in the positive direction of (OX) axis in a uniform fluid with density $\rho_2$ and flow velocity $u_2$. Along characteristics

$$\frac{x}{t} = u + c, \tag{A7}$$

where t is time, invariant $J_-$ is constant [27]. The condition of constant $J_-$ relates the uniform states ($\rho_2$, $u_2$) before and ($\rho_1$, $u_1$) behind the rarefaction fan as

$$\ln \frac{\rho_2}{\rho_1} = \frac{u_2 - u_1}{c}. \tag{A8}$$

The rarefaction fan is bounded by characteristics (A7) corresponding to flow velocities $u_1$ and $u_2$,

$$u_1 + c < \frac{x}{t} < u_2 + c. \tag{A9}$$

According to Eq. (B7), the flow velocity is a linear function of ratio *x/t* in interval (A9). Substiturion of invariant $J_-$ given by Eq. (A6) into Eq. (A7) results in the following expression for density $\rho$ in the rarefaction fan:

$$\ln \frac{\rho}{\rho_1} = \frac{x}{ct} - \frac{u_1}{c} - 1. \tag{A10}$$

*A3. Algorithm for the Riemann solver*

Consider an initial jump in plane $x = 0$ perpendicular to axis (OX) with arbitrary values of density $\rho_L$, longitudinal $x$-component of velocity $u_L$, transversal velocity component $\tau_L$, and internal energy $U_L$ to the left, $x < 0$, and the corresponding values designated by index R to the right, $x > 0$. Generally, the initial jump splits into three disturbances separated with domains of constant flow parameters. The disturbances are a contact discontinuity, up to two shocks, and up to two rarefaction fans [28]. On the contact discontinuity, normal velocity $u$ and pressure $p$ are continuous while transversal velocity $\tau$ and internal energy $U$ can be discontinuous. In the considered equation of state (24), pressure is a function of density and independent of temperature. Thus, the medium is barotropic where there is no density jump on the contact discontinuity. On the shock front, all the fluid-dynamic parameters excluding transversal velocity $\tau$ are generally discontinuous. In a rarefaction fan, an adiabatic flow is formed with continuously varying longitudinal velocity, density, pressure, and internal energy and a constant transversal velocity.

Solutions of the Riemann problem are generally classified into the following four groups according to the number and propagation direction of the arising shocks and rarefaction fans [28]: I – two shocks, one propagating to the left and another to the right; II – a rarefaction fan to the left and a shock to the right; III – a shock to the left and a rarefaction fan to the right; IV – two rarefaction fans, one to the left and another to the right. Figure A1 shows velocity profiles corresponding to the above-listed cases. In a barotropic fluid, the classification can be drawn in the phase space of the left and right initial density values and the difference of the left and right initial velocity values, $\rho_L$, $\rho_R$ and $u_R - u_L$, respectively. According to the Galilean invariance, not the left and right velocities but only their difference is an independent variable. The mentioned four cases are considered below.

*I. Two shocks, one propagating to the left and another to the right*

The Rankine-Hugoniot conditions (A1)-(A3) are written in the frame moving with the shock front. In the laboratory frame where the front moves with velocity $v$, Eqs. (A1) and (A2) become

$$\rho_1(u_1 - v) = \rho_2(u_2 - v), \tag{A11}$$

$$\rho_1(u_1 - v)^2 + p_1 = \rho_2(u_2 - v)^2 + p_2. \tag{A12}$$

One can exclude front velocity $v$ from these equations to obtain the following relation:

$$\left(\frac{1}{\rho_1} - \frac{1}{\rho_2}\right)(p_2 - p_1) = (u_2 - u_1)^2. \tag{A13}$$

Substitution of the equation of state (24) into Eq. (A13) results

$$\frac{(\rho_2 - \rho_1)^2}{\rho_1 \rho_2} = \frac{(u_2 - u_1)^2}{c^2}, \tag{A14}$$

where sound speed $c$ is given by Eq. (26).

To find constant density $\rho_M$ and longitudinal flow velocity $u_M$ between the shocks, Eq. (A14) is applied to the shocks between left L and intermediate M states and between intermediate M and right R states as follows:

$$\frac{(\rho_L - \rho_M)^2}{\rho_L \rho_M} = \frac{(u_L - u_M)^2}{c^2}, \tag{A15}$$

$$\frac{(\rho_R - \rho_M)^2}{\rho_R \rho_M} = \frac{(u_R - u_M)^2}{c^2}. \tag{A16}$$

Notice that in the considered case I, the following inequalities are valid:

$$\rho_L < \rho_M > \rho_R, \quad u_L > u_M > u_R. \tag{A17}$$

Therefore, Eqs. (A15) and (A16) can be written as

$$\frac{\rho_M - \rho_L}{\sqrt{\rho_L \rho_M}} = \frac{u_L - u_M}{c}, \tag{A18}$$

$$\frac{\rho_M - \rho_R}{\sqrt{\rho_R \rho_M}} = \frac{u_M - u_R}{c}. \tag{A19}$$

Addition of Eqs. (A18) and (A19) excludes $u_M$,

$$(s_L + s_R)(s_M^2 - s_L s_R) = \omega s_L s_M s_R, \tag{A20}$$

where $\omega = (u_L - u_R)/c$ is the dimensionless velocity jump and $s = \rho^{1/2}$.

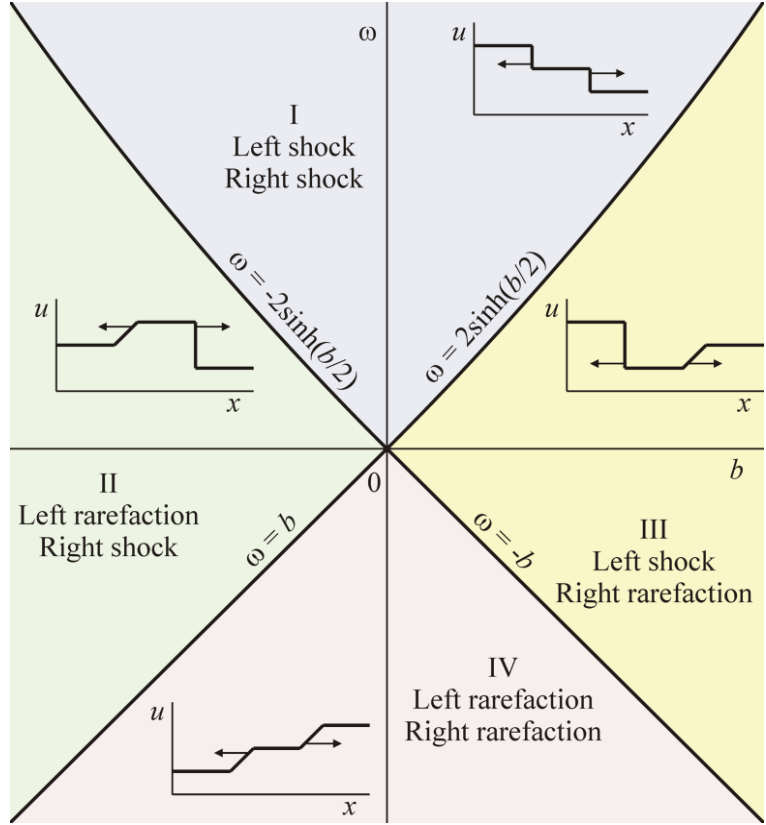

**Fig. A1.** Four possible cases for the Riemann problem solution in the phase plane of initial states ($b$-$\omega$).

In case I, intermediate density $\rho_M$ is always greater than left $\rho_L$ and right $\rho_R$ densities, see Eq. (A17). If $\rho_M$ decreases and approaches $\rho_L$, the left shock vanishes. When $\rho_M$ becomes less than $\rho_L$, the left shock becomes a rarefaction fan. Therefore, condition

$$\rho_L = \rho_M, \tag{A21}$$

is the boundary between cases I and II. Substituting Eq. (A21) into Eq. (A20) reduces this condition to

$$\text{I/II:} \quad \sinh\frac{b}{2} = -\frac{\omega}{2}, \tag{A22}$$

where $b = \ln(\rho_R/\rho_L)$ is the dimensionless density jump. If $\rho_M$ decreases and approaches $\rho_R$, the right shock vanishes. When $\rho_M$ becomes less than $\rho_R$, the right shock becomes a rarefaction fan. Similarly, condition

$$\rho_R = \rho_M, \tag{A23}$$

is the boundary between cases I and III. Substituting Eq. (A23) into Eq. (A20) reduces this condition to

$$\text{I/III:} \quad \sinh\frac{b}{2} = \frac{\omega}{2}. \tag{A24}$$

Fig. A1 shows curves (A22) and (A24) in the phase plane of initial states ($b$-$\omega$).

The density in the intermediate state can be found from Eq. (A20). This is a quadratic equation relative $s_M$ with the unique non-negative root

$$s_M = a + \sqrt{a^2 + s_L s_R}, \qquad a = \frac{\omega s_L s_R}{2(s_L + s_R)}. \tag{A25}$$

The density itself is $\rho_M = s_M^2$. Then, longitudinal velocity $u_M$ in the intermediate state can be found from Eqs. (A18) or (A19). The above found values of $\rho_M$ and $u_M$ are the same to the both sides of the contact discontinuity, while the values of internal energy $U$ and transversal component of flow velocity $\tau$ can be different. The values of transversal velocity to the left and right of the contact discontinuity are left $\tau_L$ and right $\tau_R$ initial values, respectively. Therefore, intermediate value of transversal velocity $\tau_M$ at the place of the initial jump depends on the sign of $u_M$,

$$\tau_M = \begin{cases} \tau_L, & u_M > 0 \\ \tau_R, & u_M < 0 \end{cases}. \tag{A26}$$

The values of internal energy to the left and right of the contact discontinuity are related to the initial values to the left $U_L$ and right $U_R$ by the Hugoniot adiabat (A5). Intermediate value of enthalpy $H_M$ at the place of initial jump depends on the sign of $u_M$,

$$\frac{H_M}{\rho_M} = \begin{cases} \dfrac{H_L}{\rho_L} + \dfrac{c^2}{2}\left(\dfrac{\rho_M}{\rho_L} - \dfrac{\rho_L}{\rho_M}\right), & u_M > 0 \\ \dfrac{H_R}{\rho_R} + \dfrac{c^2}{2}\left(\dfrac{\rho_M}{\rho_R} - \dfrac{\rho_R}{\rho_M}\right), & u_M < 0 \end{cases}, \tag{A27}$$

where Eq. (A4) defines the relation between internal energy $U$ and enthalpy $H$.

*II. A rarefaction fan to the left and a shock to the right*

Equation (A8) can be applied to the left rarefaction fan and becomes

$$\ln\frac{\rho_L}{\rho_M} = \frac{u_M - u_L}{c}. \tag{A28}$$

Equation (A19) on the right shock is still applicable. Excluding $u_M$ from Eqs. (A19) and (A28) gives

$$\frac{\rho_M - \rho_R}{\sqrt{\rho_R \rho_M}} + \ln\frac{\rho_M}{\rho_L} = \omega. \tag{A29}$$

If $\rho_M$ decreases and approaches $\rho_R$, the left shock vanishes. When $\rho_M$ becomes less than $\rho_R$, the right shock becomes a rarefaction fan. Currently, condition (A23) is the boundary between cases II and IV. Substituting Eq. (A23) into Eq. (A29) results in

$$\text{II/IV:} \quad b = \omega. \tag{A30}$$

Boundary (A30) is shown on the phase plane of initial states ($b$-$\omega$) in Fig. A1.

The value of density $\rho_M$ in the intermediate state is the solution of transcendent equation (A29). One can substitute this solution into Eq. (A28) to find flow velocity $u_M$ in the intermediate state. The intermediate value of transversal velocity $\tau_M$ at the place of the initial jump is always defined by Eq. (A26). Internal energy $U$ is a sum of thermal $\mathcal{T}$ and elastic $\mathcal{E}$ components according to Eq. (25). In the considered barotropic fluid, there is no interaction between the thermal and elastic energy components. Therefore, thermal energy is conserved in adiabatic flow in a rarefaction fan. The value of internal energy $U_M$ at the place of the initial jump is calculated with the thermal energy conservation in the rarefaction fan or Hugoniot adiabat for the shock,

$$\begin{cases} \dfrac{\mathcal{T}_M}{\rho_M} = \dfrac{\mathcal{T}_L}{\rho_L} & , \quad u_M > 0 \\ \dfrac{H_M}{\rho_M} = \dfrac{H_R}{\rho_R} + \dfrac{c^2}{2}\left(\dfrac{\rho_M}{\rho_R} - \dfrac{\rho_R}{\rho_M}\right) & , \quad u_M < 0 \end{cases}. \tag{A31}$$

*III. A shock to the left and a rarefaction fan to the right*

Equation (A8) can be applied to the right rarefaction fan and becomes

$$\ln\frac{\rho_R}{\rho_M} = \frac{u_R - u_M}{c}. \tag{A32}$$

One can apply Eq. (A18) on the left shock as well. Excluding $u_M$ from Eqs. (A18) and (A32) results

$$\frac{\rho_M - \rho_L}{\sqrt{\rho_L \rho_M}} + \ln\frac{\rho_M}{\rho_R} = \omega. \tag{A33}$$

If $\rho_M$ decreases and approaches $\rho_L$, the left shock vanishes. When $\rho_M$ becomes below $\rho_L$, the left shock becomes a rarefaction fan. Therefore, condition (A21) is the boundary between cases III and IV. Substituting Eq. (A21) into Eq. (A33) results

$$\text{III/IV:} \quad b = -\omega. \tag{A34}$$

Boundary (A34) is shown in the phase plane of initial states ($b$-$\omega$) in Fig. A1.

The value of density $\rho_M$ in the intermediate state is the solution of transcendent equation (A33). This solution is substituted into Eq. (A32) to find the value of longitudinal velocity $u_M$ in the intermediate state. The value of transversal velocity $\tau_M$ at the place of the initial jump is given by Eq. (A26). The value of internal energy $U_M$ at the place of the initial jump is calculated with the conservation of thermal energy in the rarefaction fan or Hugoniot adiabat on the shock,

$$\begin{cases} \dfrac{H_M}{\rho_M} = \dfrac{H_L}{\rho_L} + \dfrac{c^2}{2}\left(\dfrac{\rho_M}{\rho_L} - \dfrac{\rho_L}{\rho_M}\right) &, \quad u_M > 0 \\ \dfrac{\mathcal{T}_M}{\rho_M} = \dfrac{\mathcal{T}_R}{\rho_R} &, \quad u_M < 0 \end{cases}. \quad (A35)$$

*IV. Two rarefaction fans, one to the left and another to the right*

In case IV, conditions (A28) and (A32) are valid in the left and right rarefaction fans, respectively. These equations have explicit solution

$$\rho_M = \sqrt{\rho_L \rho_R}\,\exp\left(\dfrac{\omega}{2}\right), \quad u_M = \dfrac{u_L + u_R}{2} - \dfrac{bc}{2}. \quad (A36)$$

The value of transversal velocity $\tau_M$ at the place of initial jump is given by Eq. (A26). The value of internal energy $U_M$ at the place of the initial jump is calculated with thermal energy conservation in a rarefaction fan,

$$\dfrac{\mathcal{T}_M}{\rho_M} = \begin{cases} \dfrac{\mathcal{T}_L}{\rho_L} &, \quad u_M > 0 \\ \dfrac{\mathcal{T}_R}{\rho_R} &, \quad u_M < 0 \end{cases}. \quad (A37)$$

According to the above Riemann problem solution, a flow chart is developed and shown in Fig. A2. The algorithm includes the classification of solutions based on the input data and calculations for each of the four possible cases. The output data are the parameters in the intermediate state at the place of the initial jump. In this algorithm, it is assumed that the absolute value of longitudinal flow velocity in the intermediate state $u_M$ is less than sound speed $c$. Thus, at the place of the initial jump, there cannot be a rarefaction fan or initial states on the left or right. Such a simplification is acceptable for subsonic flows.

**Appendix B: Melt pool parameters calculated with the Rosenthal model**

Table B1 lists numerical results obtained with the algorithms described in Section 2.1.

**Table B1.** Rosenthal models: Dimensionless parameters of the melt pool

| Π | 1/32 | 1/16 | 1/8 | 1/4 | 1/2 | 1 | 2 | 4 | 8 | 16 | 32 | 64 | 128 |
|---|---|---|---|---|---|---|---|---|---|---|---|---|---|
| $r_-/L_0$ | .031758 | .11522 | .23372 | .30241 | .29320 | .23757 | .17176 | .11514 | .07324 | .04486 | .02670 | .01555 | .00890 |
| $r_+/L_0$ | .030153 | .12420 | .29537 | .49701 | .67591 | .80746 | .89240 | .94244 | .97008 | .98472 | .99228 | .99612 | .99805 |
| $L/L_0$ | .061911 | .23942 | .52909 | .79942 | .96912 | 1.0450 | 1.0642 | 1.0576 | 1.0433 | 1.0296 | 1.0190 | 1.0117 | 1.0070 |
| $r_m/L_0$ | .029954 | .11964 | .26322 | .39174 | .46079 | .47568 | .46021 | .43471 | .41084 | .39314 | .38180 | .37523 | .37167 |
| $x_m/L_0$ | .00020 | .00449 | .03065 | .09525 | .18196 | .26039 | .31462 | .34512 | .35944 | .36512 | .36707 | .36766 | .3678 |
| $D/L_0$ | .029953 | .11956 | .26143 | .37998 | .42334 | .39808 | .33587 | .26432 | .19899 | .14576 | .10505 | .075032 | .053329 |
| $D/D_0$ | .0087300 | .049280 | .15239 | .31324 | .49354 | .65633 | .78314 | .87158 | .92794 | .96125 | .97978 | .98965 | .99476 |
| $L/D$ | 2.0669 | 2.0025 | 2.0238 | 2.1038 | 2.2892 | 2.6251 | 3.1683 | 4.0011 | 5.2431 | 7.0637 | 9.6997 | 13.483 | 18.881 |

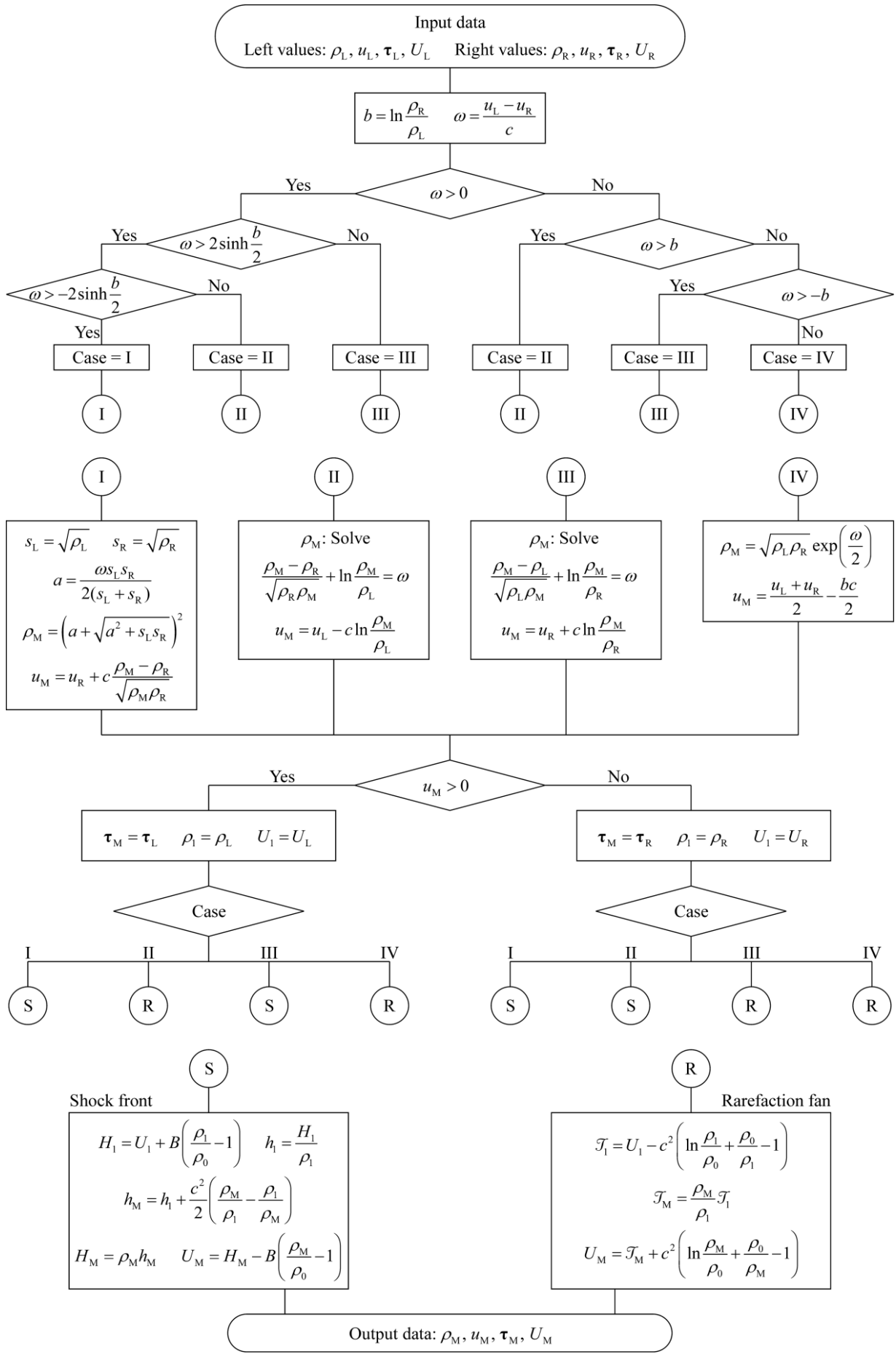

**Fig. A2.** Algorithm of the Riemann solver.